\documentclass[twocolumn,aps,showpacs,prb]{revtex4-1}
\usepackage{amsmath,amssymb}
\usepackage{subfig}
\usepackage{graphicx}
\graphicspath{{figures/}}
\usepackage{braket}
\usepackage{epstopdf}
\usepackage[usenames]{color}
\usepackage{hyperref}
\usepackage{float}

\begin{document}

\title{Criterion for the occurrence of many body localization in the presence of a single particle mobility edge}

\author{Ranjan Modak} 
\author{Soumi Ghosh}
\author{Subroto Mukerjee }
\affiliation{Department of Physics, Indian Institute of Science, Bangalore 560 012, India}


\begin{abstract}
Non-interacting fermions in one dimension can undergo a localization-delocalization transition in the presence of a quasi-periodic potential as a function of that potential.
In the presence of interactions, this transition transforms into a Many-Body Localization (MBL) transition.
Recent studies have suggested that this type of transition can also occur in models with quasi-periodic potentials that possess single particle mobility edges.
Two such models were studied in PRL 115,230401(2015) but only one was found to exhibit an MBL transition in the presence of interactions while the other one did not. In this work we investigate the occurrence of MBL in the presence of weak interactions in five different models with single particle mobility edges in one dimension with a view to obtaining a criterion for the same. We find that not all such models undergo a thermal-MBL phase transition in presence of weak interactions.
We propose a criterion to determine whether MBL is likely to occur in presence of interaction based only on the properties of the non-interacting models.
The relevant quantity $\epsilon$ is a measure of how localized the localized states are relative to how delocalized the delocalized states are in the non-interacting model.
We also study various other features of the non-interacting models such as the divergence of the localization length at the mobility edge and the presence or absence of `ergodicity' and localization in their many-body eigenstates. However, we find that these features cannot be used to predict the occurrence of MBL upon the introduction of weak interactions. 
\end{abstract}


\maketitle
\section{Introduction}
 An arbitrarily weak amount of disorder can localize all eigenstates in a non-interacting quantum system in dimensions $d \le 2$ while in three dimensions a mobility edge of energy $E_c$ can exist that separates localized and delocalized states~\cite{anderson.1958,tvr.1979,tvr.1985}.  In the presence of interactions, for large enough disorder, a many-body version of this effect called Many-Body Localization(MBL) can occur~\cite{basko.2006}, which has attracted a lot of attention recently~\cite{serbyn.2013,gopalakrishnan.2015,agarwal.2014,bardarson.2012}. Systems displaying MBL are the only known examples of generic interacting isolated quantum systems that do not thermalize~\cite{vadim.2007,pal.2010} and thus do not obey the Eigenstate Thermalization Hypothesis (ETH)~\cite{deutsch1991,srednicki1994,rigol.2008} . It has been argued these systems possess emergent conservation laws~\cite{huse2014phenomenology,modak1.2015,serbyn2013local,chandran2015constructing} which prevent thermalization like in integrable systems~\cite{rigol1.2009,santos.2010}. Systems which display MBL typically undergo  a thermal-MBL transition  as a function of quenched disorder.

A delocalization-localization transition analogous to the thermal-MBL transition can occur even in a non-interacting system. An example of a one-dimensional microscopic model with such a transition is the Aubry-Andre model (AA model)~\cite{aubry.1980}. The transition is between a phase with {\em all} single particle states localized and one with {\em all} states delocalized. The delocalization-localization transition in this model transforms into a thermal-MBL transition upon the introduction of interactions~\cite{huse.2013}. The AA model has also been emulated in experiments on cold-atoms in the non-interacting limit~\cite{fallani.2007,lucioni.2011} and with interactions to observe MBL~\cite{ehud.2015}. It has also been proposed that the thermal-MBL transition in the AA model in the presence of interactions is in a different universality class from the transition in models with true disorder~\cite{uniclass.2017}.

The AA model can be modified to yield single-particle mobility edges~\cite{griniasty.1988,dassarma.1990,ganeshan.2014} and one such model has also been recently realized experimentally~\cite{spme_expt.2017}. Such mobility edges can also be seen in models with correlated disorder~\cite{de1998delocalization}. It was argued by Nandkishore and Potter that a non-interacting system with coexisting localized and {\em protected} delocalized states will thermalize upon the introduction of weak interactions if $\nu d \ge 1$~\cite{nandkishore2014marginal}. $d$ here is the number of physical dimensions and the localization length $\xi$ diverges at an energy $E_c$ as $\xi \sim |E-E_c|^{-\nu}$. A question that arises is whether MBL could be present in a system with localized and {\em unprotected} delocalized states upon the introduction of weak interactions. Such a situation is {\em generic} in disordered three dimensional systems and thus, the above question is of relevance to physically relevant interacting solid-state systems. Recent works~\cite{modak.2015,li2015energy,nandkishore2015many,ann_rev.2017} have shown that MBL can indeed occur in models with single particle mobility edges in one dimension. However, it was also observed that not all such models with unprotected delocalized states display MBL upon the introduction of weak interactions~\cite{modak.2015,li2015energy}. It is thus natural to ask if there is a criterion that can be formulated by an examination of the single particle spectrum of the non-interacting model to determine whether MBL can develop upon the introduction of weak interactions. The identification of such a criterion would be significant because the required calculations would have to be performed only on non-interacting models (and not interacting ones) which are possible numerically for fairly large system sizes and in some cases even analytically.

In this work, based on numerical exact diagonalization studies of several different models, we propose a criterion to determine whether a system with a single particle mobility edge displays MBL upon turning on weak interactions. Specifically, we argue that the criterion is based on a dimensionless quantity is $\epsilon$, which can be calculated from the non-interacting spectrum of a system of size $L$. It is defined as   
\begin{equation}
\epsilon= \frac{\eta(1-MPR_D/L)}{(MPR_L -1)},
\label{Eq:defratio}
\end{equation}
where $\eta$ is the ratio of the number of localized states to delocalized single particle states, $MPR_D$ is the mean participation ratio(PR) of the delocalized states and $MPR_L$ is the mean participation ratio(PR) of the localized states. (The PR of a normalized eigenstate $\psi$ is defined $PR_{\Psi }=1/\sum_{j}|c_j|^{4}$, where  $c_j$ is the amplitude of $\psi$ at site $j$. $PR  \sim 1$ for a localized state and is much larger (typical $ \sim L$) for a delocalized one.)  This quantity $\epsilon$ is thus a measure of the the relative strengths of the localized and delocalized states in the non-interacting model, i.e. how strongly localized the localized states are compared to how strongly delocalized the delocalized ones are. Our main result is that the system remains localized (thermalizes) upon the introduction of weak interactions when $\epsilon > (<1)$. This also serves as a criterion to detect the thermal-MBL transition in these systems, based on the properties of the non-interacting system. We also comment on whether the localization length exponent $\nu$ discussed in the previous paragraph is significant in determining whether MBL occurs in systems with mobility edges. 

Additionally, we study ``non-ergodic to ergodic'' transitions in the non-interacting limit in models with single particle mobility edges. It should be noted that there is no true ergodic phase for a non-interacting system. However, the scaling of the strength of the fluctuations of local quantities (such as particle number) with system size can change abruptly as a function of microscopic parameters in the models we study indicating a transition that resembles the ergodic to non-ergodic transition in interacting systems~\cite{Gan_non_int.2016}. We also make a distinction between this type of transition and a delocalization to localization transition. The latter type of transition is characterized by a change in the scaling of entanglement entropy from volume law to area law. The distinction between the concepts of delocalization and ergodicity (in the above sense) has also been discussed elsewhere~\cite{Gan_non_int.2016}. We show that the presence or absence of these phases in the non-interacting model cannot be used to predict whether MBL occurs upon the introduction of interactions.

\section{Models}
We study five different lattice models of spinless fermions of the general form 
\begin{equation}
H=-\sum_{ij} t_{ij} c^\dag_i c_j + \sum_i h_i n_i + V \sum_i n_i n_{i+1}, 
\label{Eq:modelsham}
\end{equation}
where $i$ and $j$ label sites on the lattice, $t_{ij}$ is the hopping between sites $i$ and $j$, $h_i$ is the onset potential at $i$ and $V$ is the interaction between fermions on nearest neighbor sites.  All five models have single particle mobility edges in the non-interacting limit $V=0$. The five models studied have the following parameters 
\begin{enumerate}

\item Model I: $t_{ij}=t$, when $i$ and $j$ are nearest neighbors and zero otherwise,  
$h_{i}=h\frac{\cos(2\pi i \alpha +\phi)}{1-\beta \cos(2\pi i \alpha +\phi) },$  
where $\alpha$ is an irrational number, $\beta \in (-1,1)$ and $\phi$ is an offset angle. The onsite potential is quasiperiodic and when $\beta=0$ and $V=0$, the model reduces to the AA model. For $V=0$ and $\beta \neq 0$, there is a mobility edge separating, localized and extended states at an energy $E$ given by
$\beta E=2sgn(h)(|t|-|h|/2)$~\cite{ganeshan.2014}. $\phi$ can be used to produce different realizations of this potential akin to different realizations of disorder in the case of a truly random potential.

\item Model II: c,  
$h_{i}=h\frac{1-\cos(2\pi i \alpha +\phi)}{1+\beta \cos(2\pi i \alpha +\phi) },$
with $\beta \in (-1,1)$, irrational $\alpha$ and offset angle $\phi$. When $V=0$, this model also reduces to the AA model for $\beta=0$ and for $\beta \neq 0$, there is a mobility edge separating, localized and extended states at an energy $E$ given by $\beta E=2sgn(h)(|t|-|h|/2)$~\cite{ganeshan.2014}.

\item Model III: $t_{ij}=t$, when $i$ and $j$ are nearest neighbors and zero otherwise,  $h_{i}=h\cos(2\pi \alpha i^{n} +\phi)$  with $0<n<1$ and irrational $\alpha$ and offset angle $\phi$. For $V=0$ and $n=1$, this is just the AA model. However, for $n<1$ and $V=0$, the model has a single-particle mobility edge when $h<2t$~\cite{griniasty.1988,dassarma.1990}. All single particle states with energy between $\pm |2t-h|$ are delocalized and all other states are localized. For $h>2t$ all single particle states are localized as in the usual AA model.

\item Model IV: $t_{ij}=t$, when $i$ and $j$ are nearest neighbors, $t_{ij}=t'$  when $i$ and $j$ are next nearest neighbors and zero otherwise. $h_{i}=h\cos(2\pi \alpha i+\phi)$ with irrational $\alpha$ and offset angle $\phi$. For $V=0$, this model too has a mobility edge separating localized and delocalized states whenever a non-zero next nearest neighbor hopping ($t'$) is present along with the nearest neighbor hopping t~\cite{biddle.2009}.

\item Model V: $t_{ij}=t$, when $i$ and $j$ are nearest neighbors and zero otherwise. $h_{i}=\sum _{k=1}^{L/2}(k^{-\gamma}(2\pi/L)^{(1-\gamma)})^{1/2}\cos(2\pi  ik/L+\phi_k)$. $\phi_k$ are $L/2$ independent random phases uniformly distributed in the interval $[0,2\pi]$ with the normalization $\sqrt{(<h_i^2>-<h_i>^2)}=1$. Unlike the other four models, this one has long-range correlated disorder as opposed to a quasi-periodic potential. It has a single particle mobility edge separating localized and delocalized states for $\gamma>2$ and  $V=0$~\cite{de1998delocalization}.
\end{enumerate}

 \begin{figure}[t]
 \centering
\includegraphics[width=2.5in]{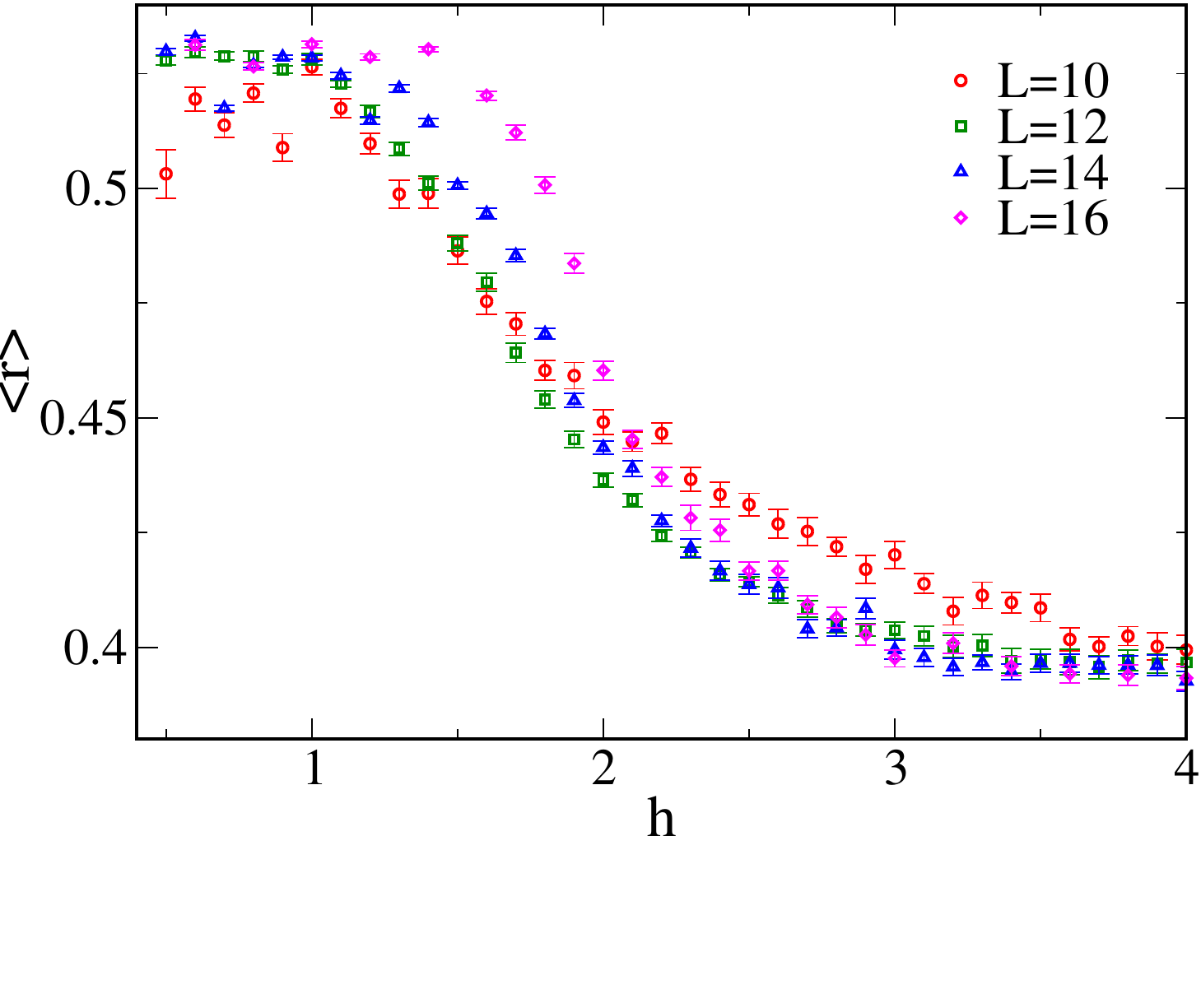} 
\qquad
\includegraphics[width=2.5in]{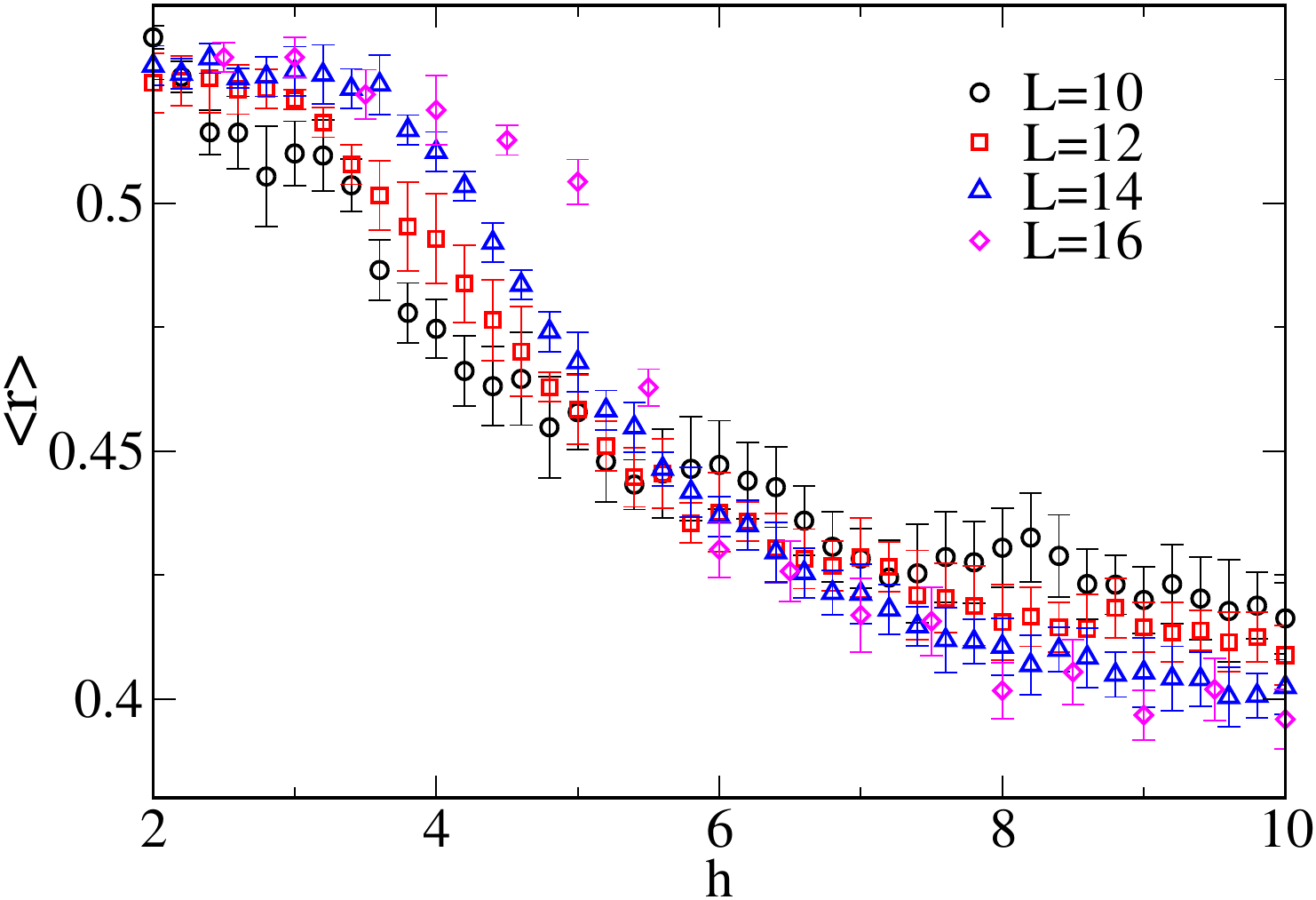} 
\caption{\small{(Color Online)The variation of the mean of the ratio between adjacent gaps in the spectrum with the strength of the incommensurate 
potential $h$ for $L=10,12,14,16$ at half filling for (top) model I ($V=1$ and $\beta=-0.6$) and (bottom) model II 
($V=1$ and $\beta=-0.75$). 
The results shown here are averaged over several realizations for flux $\phi$ which are randomly chosen from 
intervals $[0,2\pi]$ with uniform probability. The error bars denote standard deviation.
The thermal to MBL transition can be estimated from the crossing of the different curves and is at 
$h \approx 2$ for model I and $h \approx 6$ for model II. }}
\label{Fig:level spacing I}
\end{figure}

\section{Interactions}
We study the above models in the presence of interactions ($V \neq 0$) to determine whether they display MBL and to locate the resulting thermal-MBL transition. Note that in the presence of interactions, there is no apparent distinction between delocalization and ergodicity and similarly between localization and non-ergodicity. Thus, there is only one possible transition which we call a thermal-MBL transition. Our calculations employ exact diagonalization on finite-sized systems up to size $L=16$ with an average over the offset $\phi$ for better statistics. We set $t=1$ and $\alpha=\frac{\sqrt{5}-1}{2}$ and work at half filling. 
 
We first determine which of the above models display MBL and locate the thermal-MBL transition in them. We do this by studying the energy level-spacing distribution and the scaling of the entanglement entropy of mid-spectrum states as explained below.

\begin{figure}[t]
\includegraphics[width=2.5in]{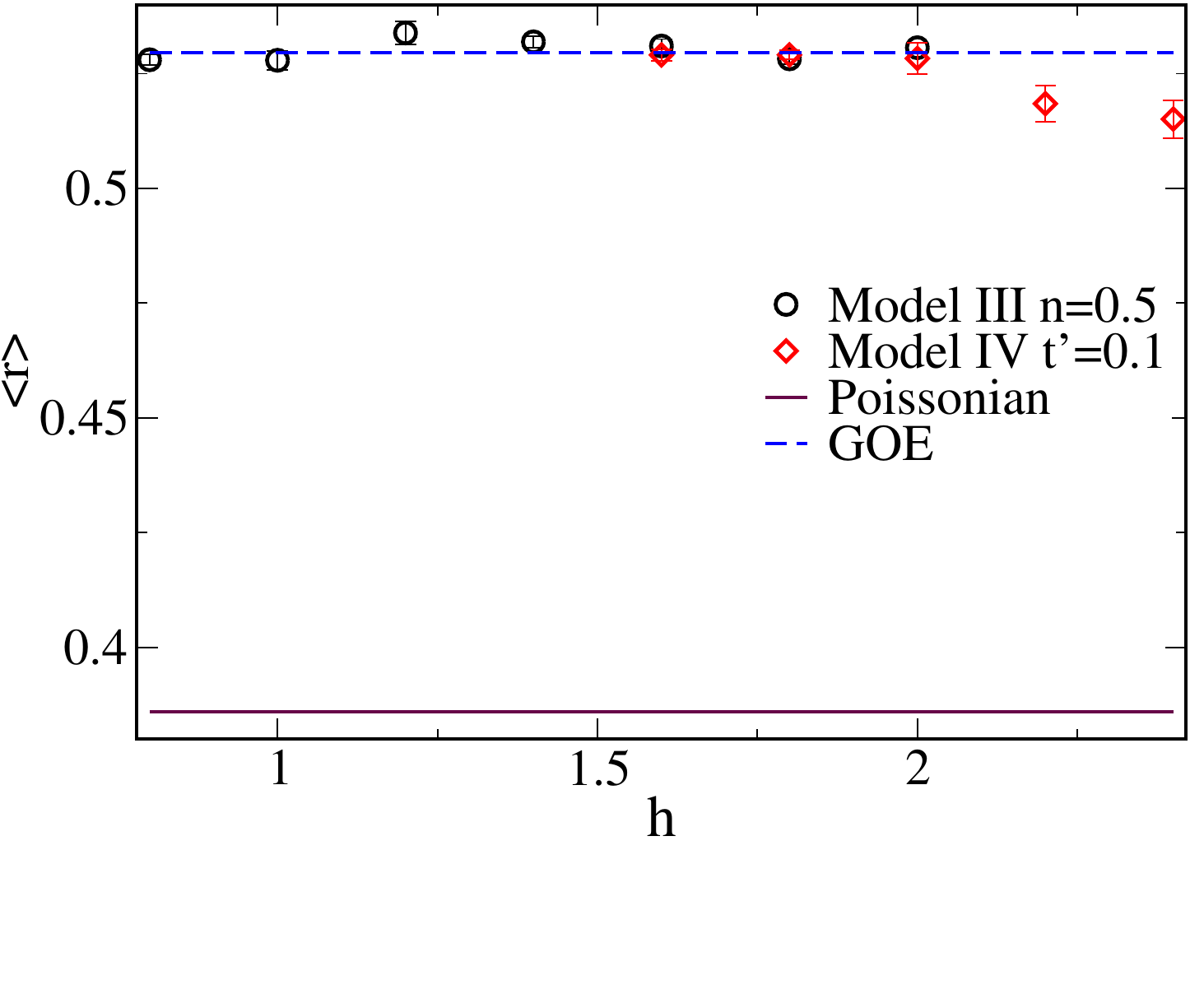} \\
\includegraphics[width=2.5in]{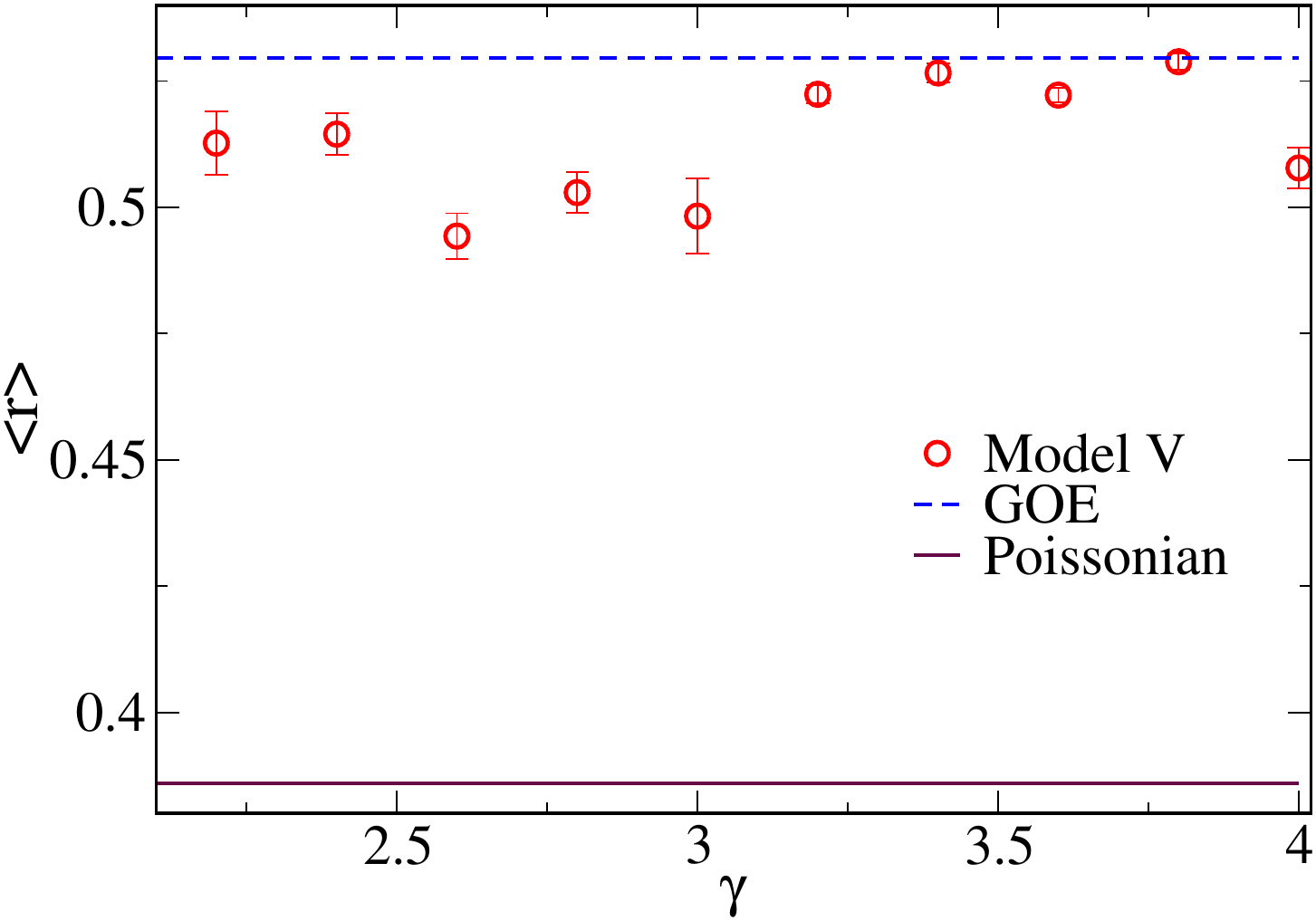} 
\caption{\small{(Color Online) The variation of the mean of the ratio between adjacent gaps in the spectrum (top) 
with the strength of the incommensurate potential $h$ for $L=16$ at half filling for  model III ($V=1$ and $n=0.5$)  
and model IV ($V=1$ and $t'=0.1$) and (bottom) with $\gamma$ for model V. 
The results shown here are averaged over several realizations for flux $\phi$ which are randomly chosen from 
intervals $[0,2\pi]$ with uniform probability. The error bars denote standard deviation.
The solid line corresponds to Poissonian and the dashed line corresponds  GOE level spacing distributions. 
It can be seen that unlike models I and II, models III, IV and V are always in the thermal phase.}}
\label{Fig:ls III}
\end{figure}
 
 \paragraph*{Energy level spacing statistics:}
Energy level spacing statistics can be used to estimate the location of the MBL transition. At the transition, the statistics change from being 
Wigner-Dyson-like specifically of the Gaussian Orthogonal Ensemble (GOE) type (characteristic of the ergodic phase) to Poissonian (characteristic of the MBL phase) and can be tracked by the ratio of successive gaps, $r_{n}=\frac{min(\delta_n,\delta_{n+1})}{max(\delta_n,\delta_{n+1})}$~\cite{vadim.2007},
where $\delta_n=E_{n+1}-E_{n}$, the difference in energy between the $n^{\rm th}$ and $n+1^{\rm st}$ energy eigenvalues. For a Poissonian distribution the mean value of $r$ is $2\ln 2-1\approx 0.386$ while for a GOE distribution, it is $\approx 0.5295$. The distribution function $P(r) \rightarrow 0$, as $r \rightarrow 0$ in the presence of level repulsion.

For model I, with $V=1$ and  $\beta=-0.6$ and model II, with $V=1$ and  $\beta=-0.75$  as $h$ is increased, the level 
spacing distribution changes from  the GOE distribution to the Poissonian distribution for system sizes $L=10,12,14,16$
as shown in Fig.~\ref{Fig:level spacing I} . The data for different system sizes cross near $h\sim 2$ for model I and 
$h\sim 6$ for model II. Hence,  $h=2$ and $h=6$ can be considered as the locations of the thermal-MBL transition for models I and II respectively. On the other hand, models III , IV and V do not exhibit a thermal-MBL transition and the level spacing distributions of these models are of the GOE type as shown in Fig.~\ref{Fig:ls III}. Even though the non-interacting versions of these models have single particle mobility edges, switching on interactions causes them to thermalize.  

\begin{figure}[t]
\includegraphics[width=2.5 in]{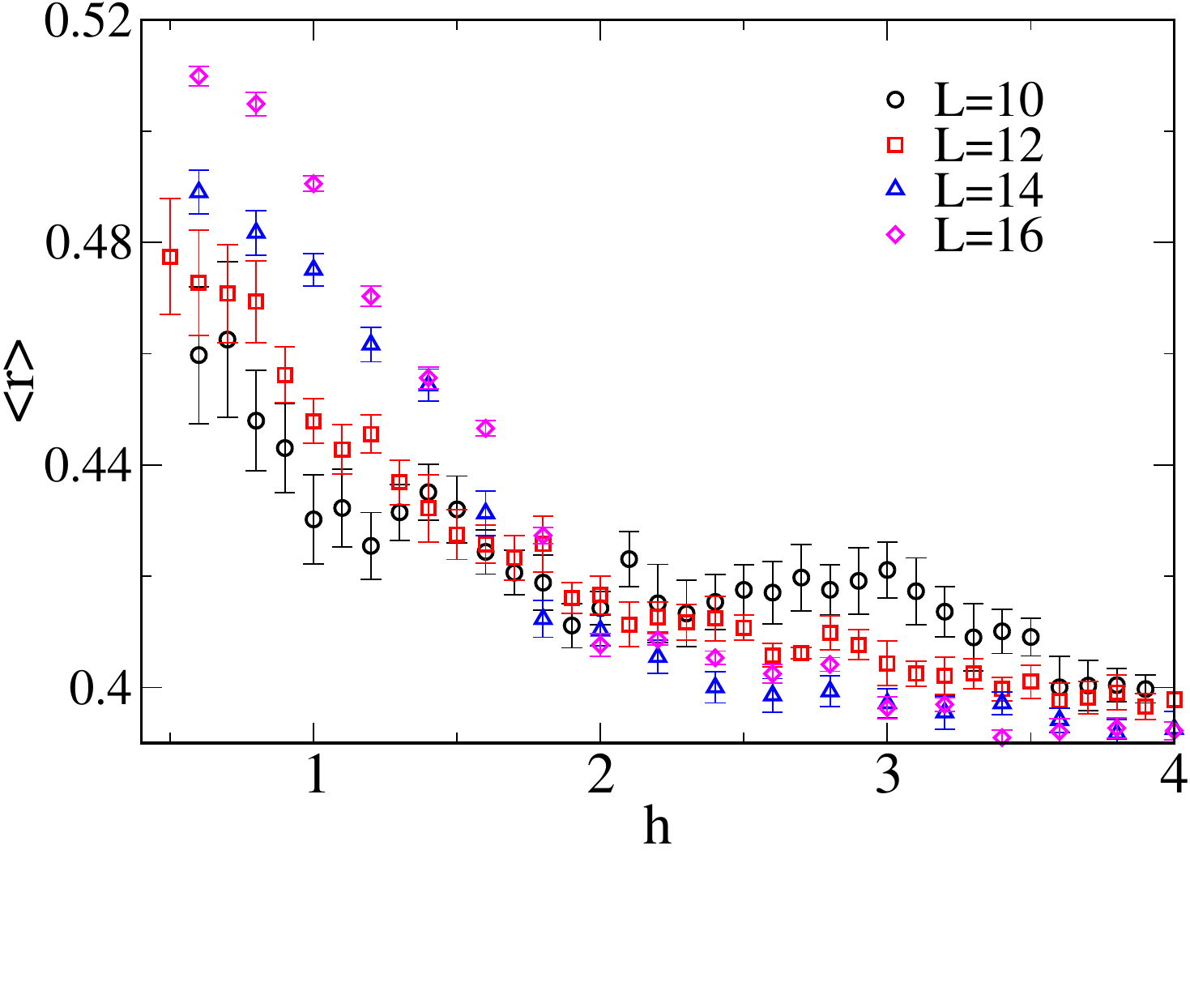} 
\includegraphics[width=2.5 in]{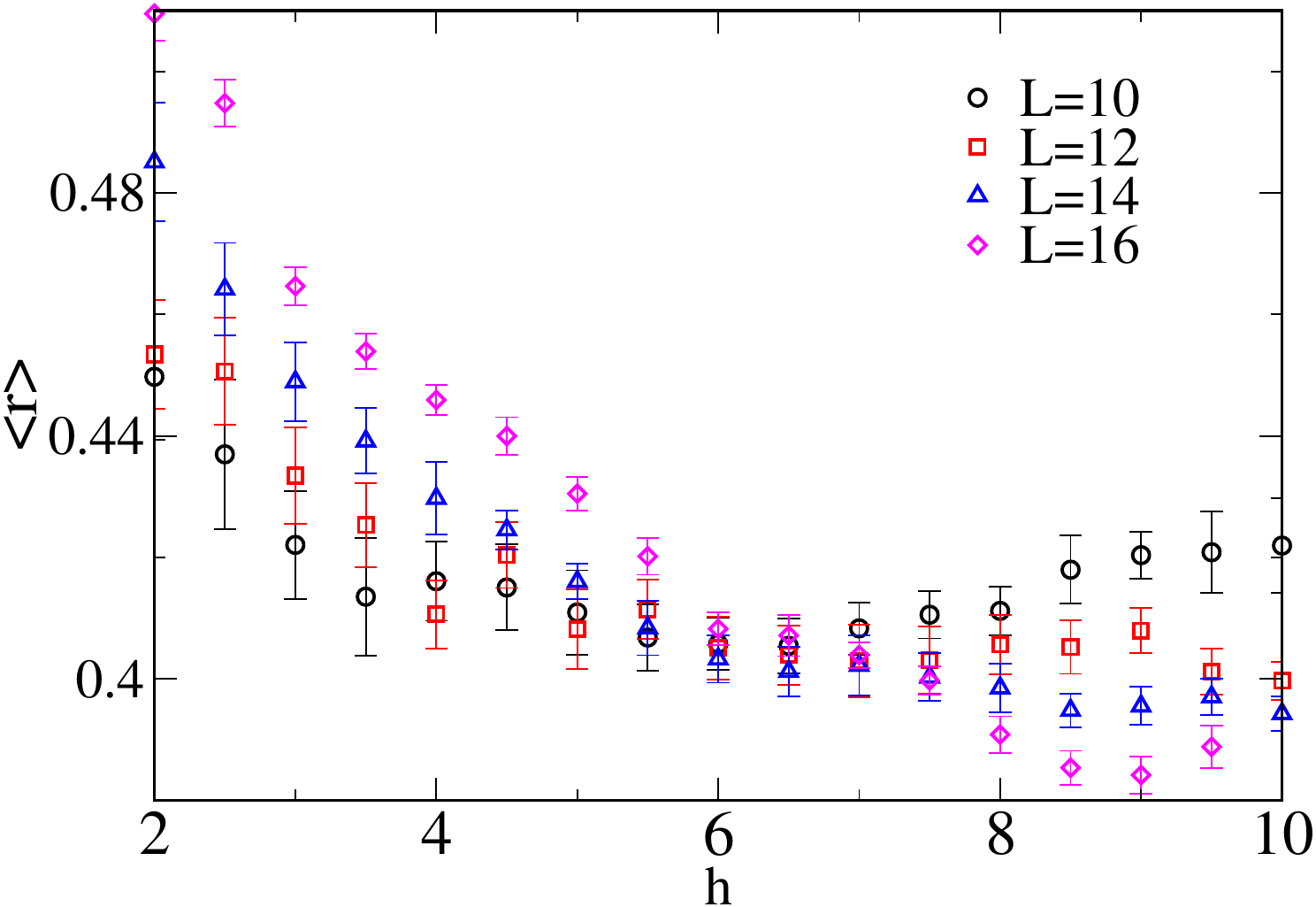}
\caption{(Color Online) The variation of the mean of the ratio between adjacent gaps in the spectrum as a function of 
the strength of the incommensurate potential $h$ for $L=10,12,14$ and $16$ at half filling for (top) 
model I ($V=0.2$ and $\beta=-0.6$) and (bottom) model II ($V=0.2$ and $\beta=-0.75$) . The results shown here are averaged over several realizations for flux $\phi$ which are randomly chosen from 
intervals $[0,2\pi]$ with uniform probability. The error bars denote standard deviation. }
\label{Fig:ls v02}
\end{figure}

\begin{figure}[t]
\includegraphics[width=2.5in]{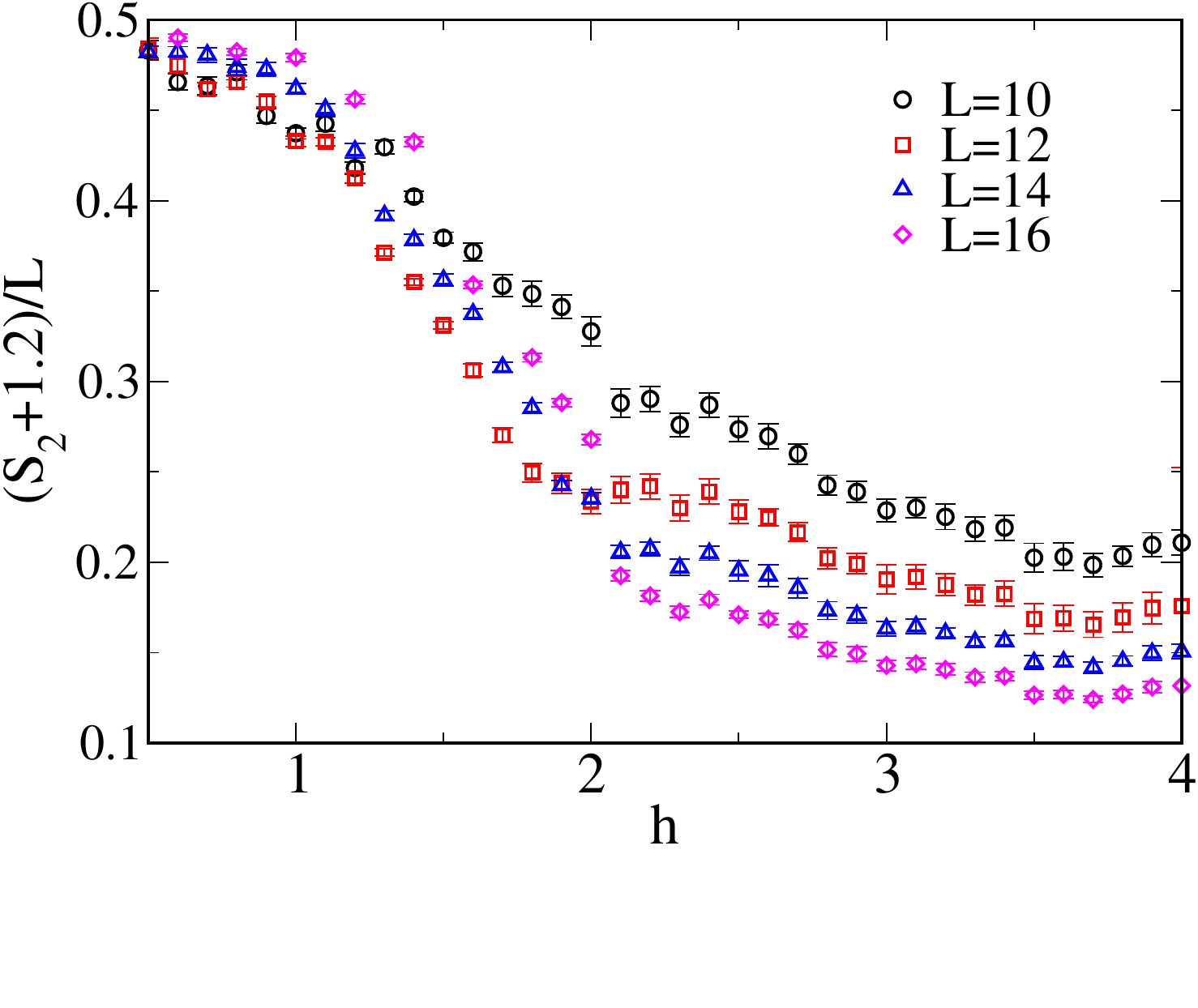} \\
\includegraphics[width=2.5in]{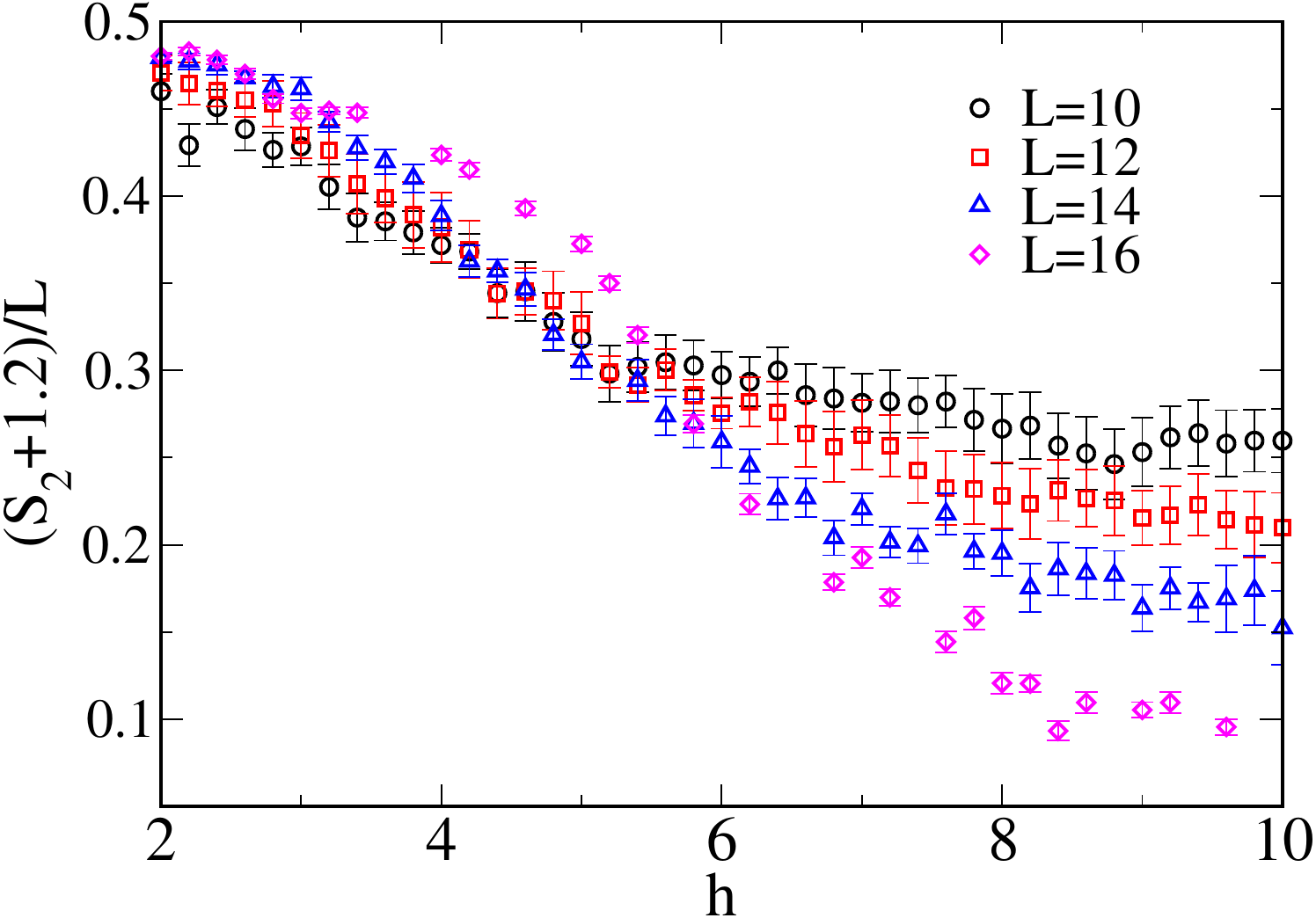}
\caption{\small{(Color Online) Variation of $(S_2+1.2)/L$ (where $S_2$ is  the Renyi entropy) of a typical mid-gap state 
for $L=10,12,14,16$ at half filling for for (top) model I ($V=1$ and $\beta=-0.6$)  and 
(bottom) model II ($V=1$ and $\beta=-0.75$). The results shown here are averaged over several realizations for flux $\phi$ which are randomly chosen from 
intervals $[0,2\pi]$ with uniform probability. The error bars denote standard deviation. The thermal-MBL transition can be estimated from the point where the curves for different values of $L$ appear to separate from one another. This is seen to be at $h \approx 2$ for model I and $h \approx 6$ for model II consistent with the values obtained from the energy level spacing distribution.}}
\label{Fig:entropy II}
\end{figure}

\begin{figure}[t]
\includegraphics[width=2.5in]{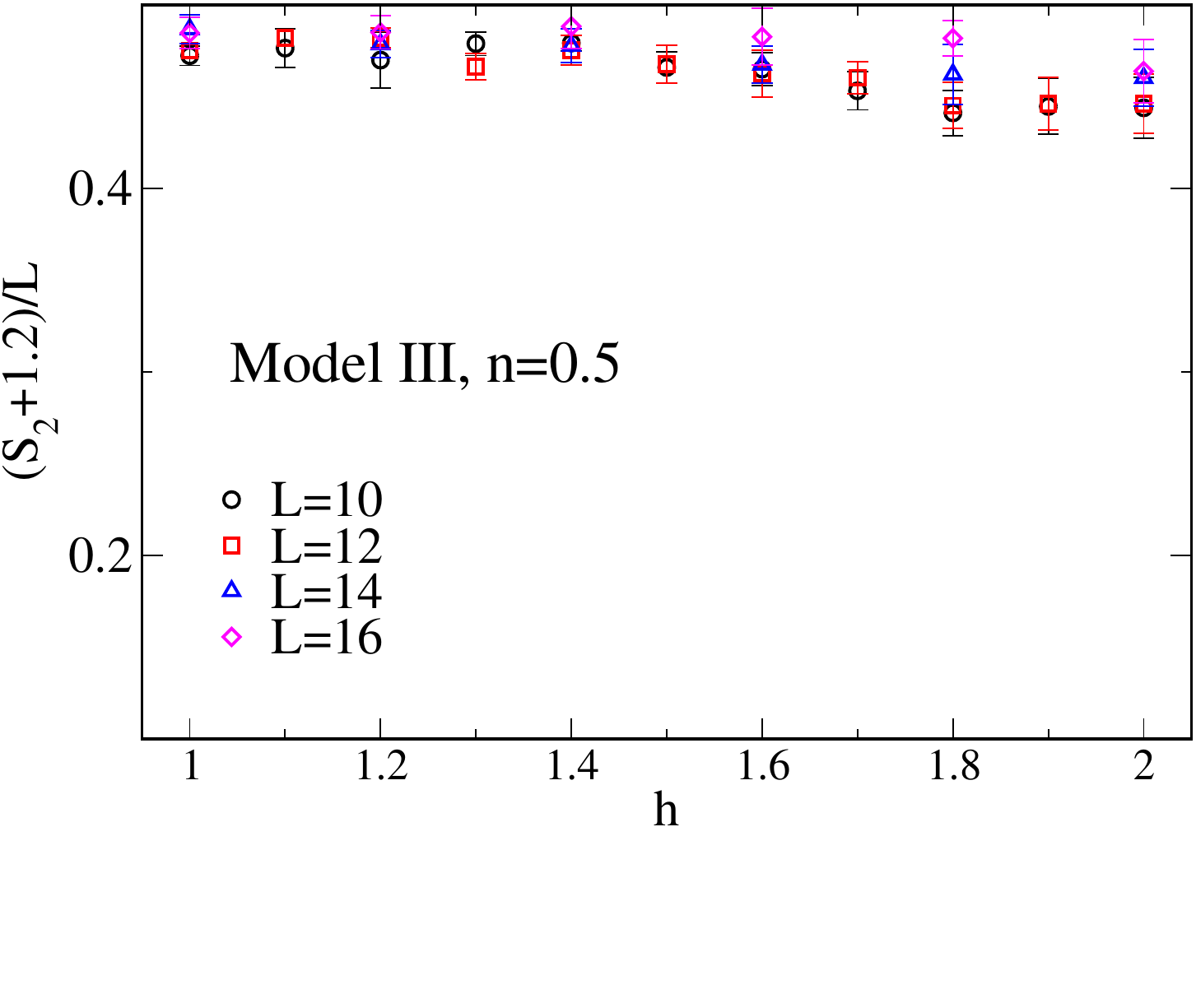} \\
\includegraphics[width=2.5in]{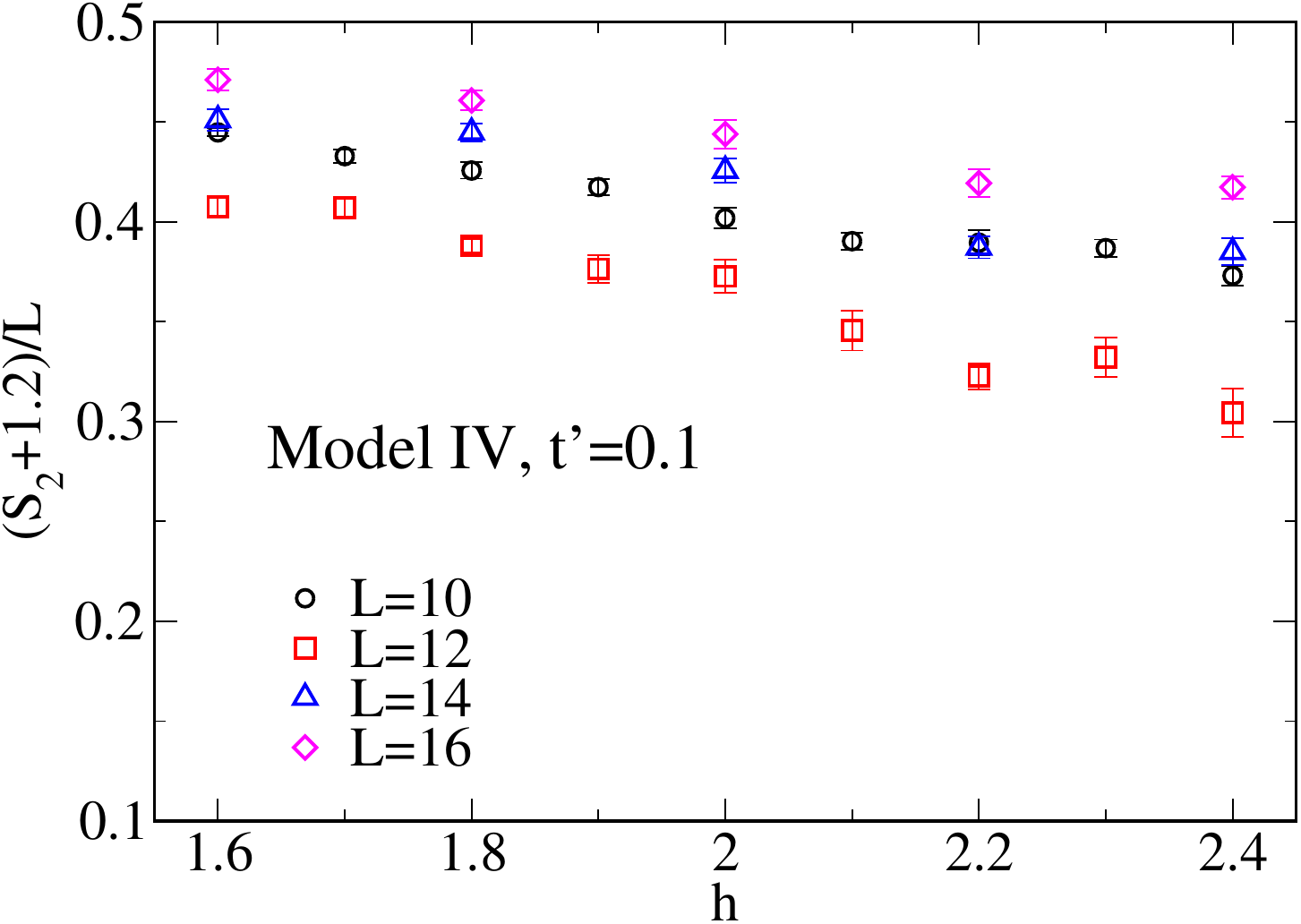}
\caption{\small{(Color Online) Variation of $(S_2+1.2)/L$ (where $S_2$ is  the Renyi entropy) of a 
typical mid-gap state for $L=10,12,14,16$ at half filling for (top) model III ($V=1$ and $n=0.5$) and 
(bottom) model IV ($V=1$ and $t'=0.1$). 
The results shown here are averaged over several realizations for flux $\phi$ which are randomly chosen from 
intervals $[0,2\pi]$ with uniform probability. The error bars denote standard deviation.
The entanglement entropy always appears to be proportional to $L$ (volume law) which indicates 
thermalization as opposed to the behavior seen for models I and II. 
This is consistent with the results obtained from the energy level spacing distribution.}}
\label{Fig:entropy III}
\end{figure}

We have shown that for Models I and II for interaction strength  $V=1.0$, there is  a thermal to Many body localization (MBL) 
transition respectively at $h\approx 2$ and $h\approx 6$. While the transition appears sharper for larger values of $V$, it also occurs for smaller values as can be seen in Fig.~\ref{Fig:ls v02} with $V=0.2$. We perform the rest of our analysis for $V=1.0$ for better clarity of numerical data.

\paragraph*{Eigenstate entanglement entropy:}
The entanglement entropy can also be used to distinguish between the thermal and many-body localized phases of a model. For a typical eigenstate (i.e. one from the middle of the spectrum), it obeys a volume law in the thermal phase and an area law in the MBL phase. We calculate the order 2 Renyi entropy $S_2=-\log_2(Tr_{A}{{\rho_{A}}^{2}})$ between the two halves $A$ and $B$ of a system of length $L$~\cite{rrnyi1961measures}. This is computationally less expensive to calculate than the von-Neumann entropy and has also recently been measured in experiments ~\cite{islam2015measuring}. $\rho_A$ is the reduced density matrix of $A$ obtained by tracing out the degrees of freedom of $B$ from the density matrix of the full system in a mid-spectrum state. For a one dimensional lattice system of spinless fermions in the ergodic phase, $S_{2}\sim \frac{L}{2}-1.2$ for system size $L$~\cite{huse.2013} and in the  many-body localized phase, $S_2\sim L^{0}$. Hence, the variation of $(S_2+1.2)/L$ with $h$ can be used as an efficient diagnostic to detect the thermal and MBL phases and the transition between them. This quantity is finite in the thermal phase and goes to zero in the MBL phase with increasing system size.

For model I with $V=1$ and  $\beta=-0.6$ and model II, with $V=1$ and  $\beta=-0.75$ as $h$ is increased $(S_2+1.2)/L$ decreases as shown in Fig.~\ref{Fig:entropy II}. From the data, the location of the thermal-MBL transition for the two models can be estimated and is consistent with their locations as obtained from energy-level spacing statistics. On the other hand, models III, IV and V do not display a thermal-MBL transition as shown in Fig.~\ref{Fig:entropy III}.

\section{Determination of the localization length exponent $\nu$}
\begin{figure}[t]
\centering
\includegraphics[width=2.0in,angle=-90]{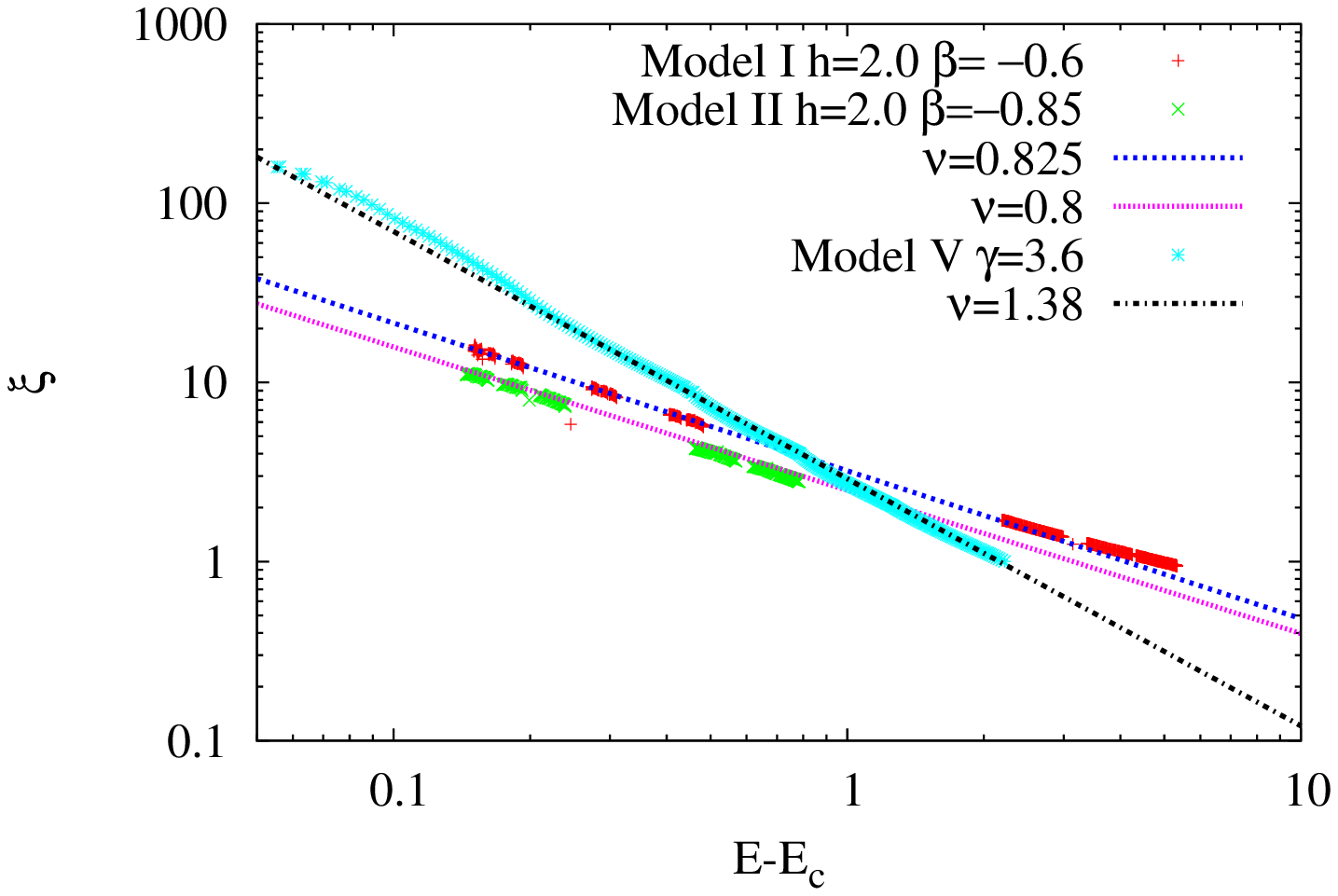} 
\includegraphics[width=2.0in,angle=-90]{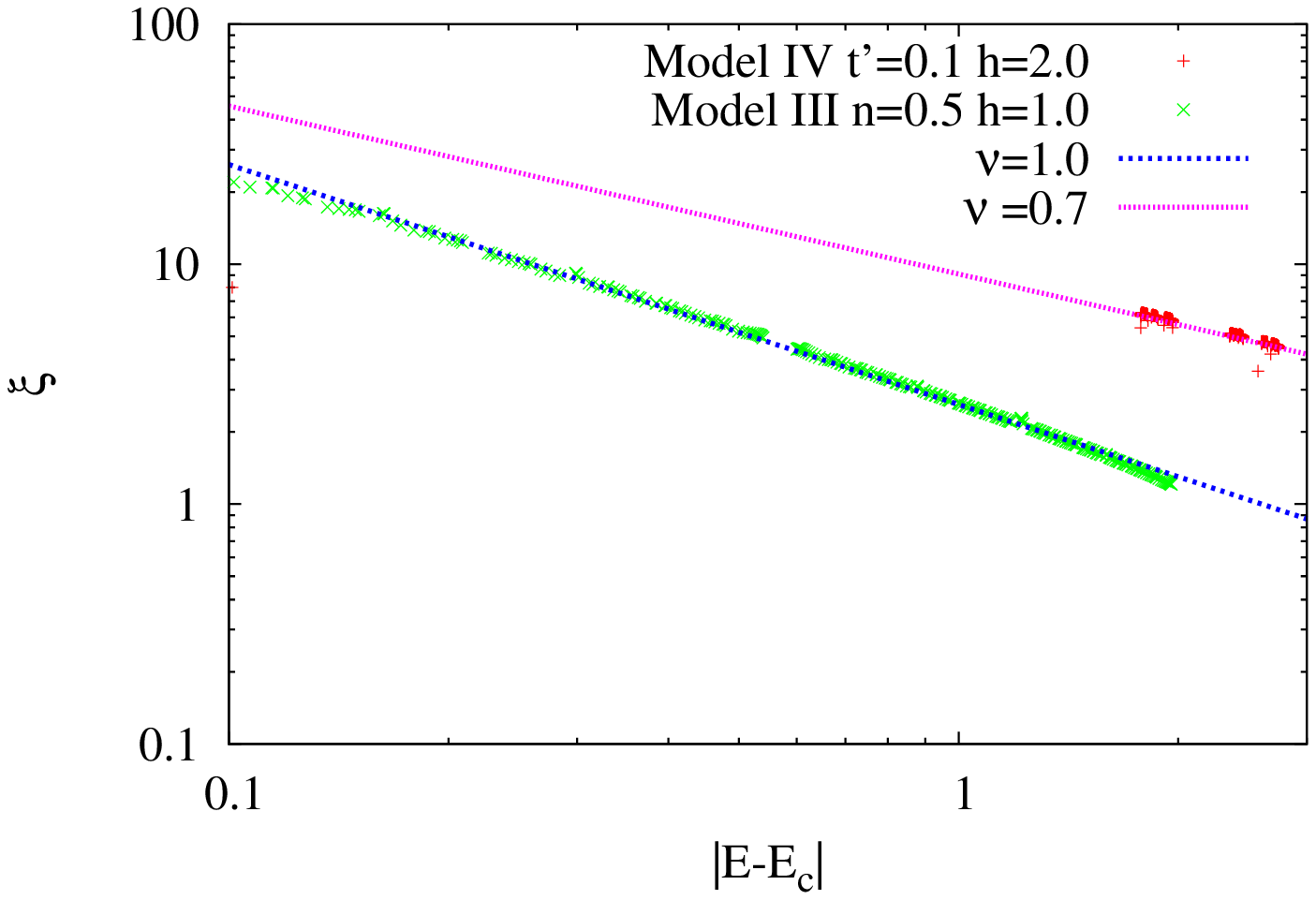} 
\caption{Localization length as function of energy for models I, II, III, IV and V }
\label{ll}
\end{figure}

The localization length of an (exponentially) localized state is finite whereas a delocalized state has infinite localization length in the thermodynamic limit.
When bands of the two types of states are separated by a mobility edge at energy $E_c$, the localization length $\xi(E)$ on the localized side typically diverges as
$\xi(E)\sim 1/|E-E_c|^{\nu}$ in the vicinity of the mobility edge. The values of $\nu$ obtained numerically for models I, II, III, IV and V  in the absence of interactions for system size $L=1000$ are shown in Fig.~\ref{ll}. The values obtained are $\nu \approx 0.825$ , $\nu \approx 0.8$, $\nu \approx 1$, $\nu \approx 0.7$ and $\nu \approx 1.38$  for models I, II, III, IV and V respectively. The values of $\nu$ for models II, III and V has also been calculated earlier~\cite{dassarma.1990,modak.2015,de1998delocalization}. Note that the values of $\nu$ for all models violate the Harris-Chayes criteria~\cite{Harris.1974} which is $\nu \geq 2/d$ where $d$ is dimension of the system. This is not surprising as the bound only applies to systems with uncorrelated disorder .

\begin{figure}[h!]
\includegraphics[width=2.5in]{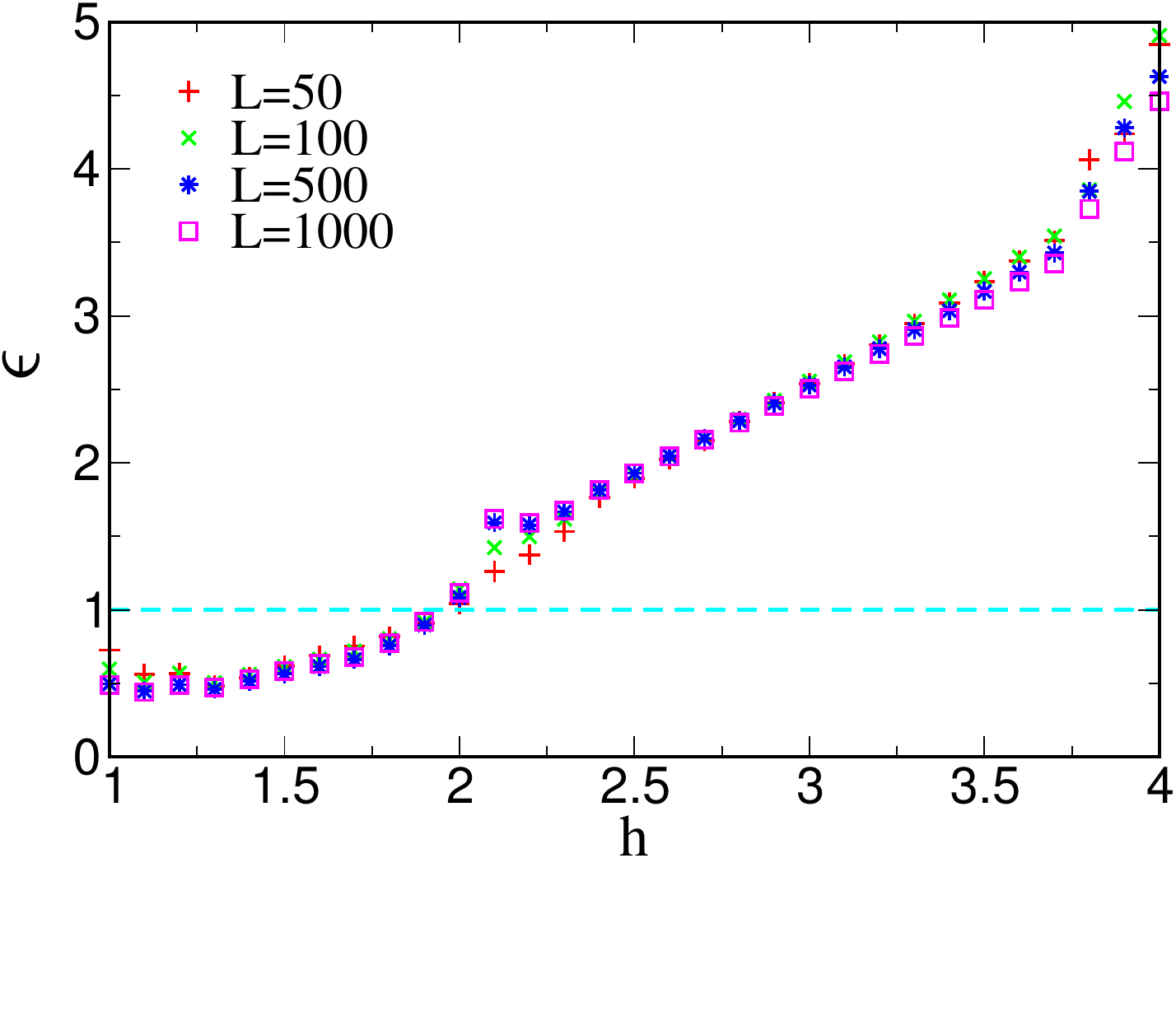} \\
\includegraphics[width=2.5in]{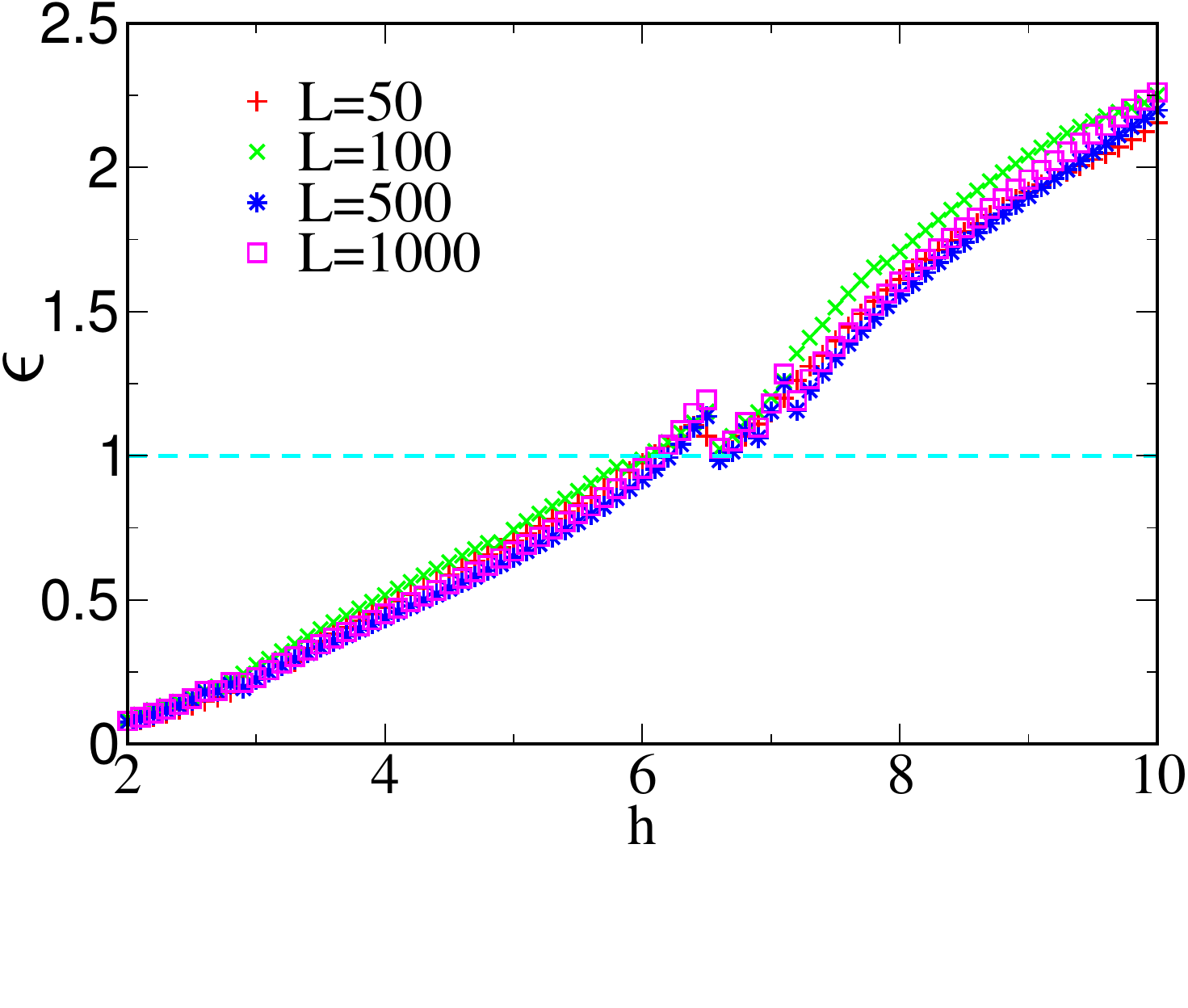} \\
\includegraphics[width=2.5in]{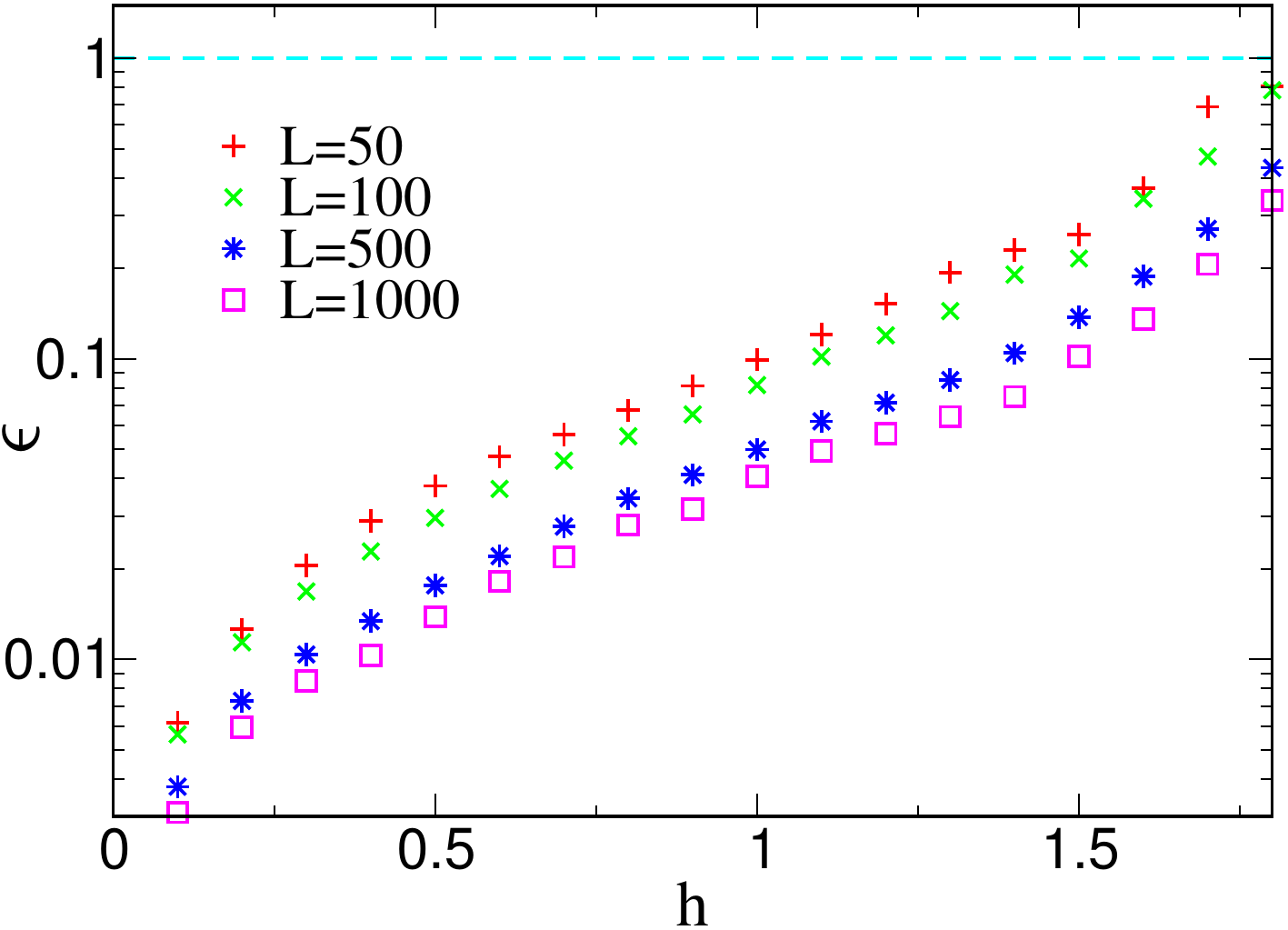} \\
\caption{\small{(Color Online) The variation of $\epsilon$ defined in Eqn.~\ref{Eq:defratio} with $h$ as obtained for $L=50,100,500,1000$  for (top) model I ( $\beta=-0.6$)  and (middle) model II (  $\beta=-0.75$) without interactions ($V=0$). The dashed lines correspond to $\epsilon=1$. The thermal-MBL transition for model I (II) has been estimated to be at $h \approx 2 (6)$ from the level spacing statistics and entanglement entropy of typical mid-spectrum states. It can be seen that $\epsilon < 1$ for $h$ below the transition value (thermal phase) and $\epsilon > 1$ for $h$ above the transition value (MBL phase). (Bottom)  $\epsilon$ as a function of $h$ for model III ($n=0.5$) for $V=0$. Model III always thermalizes upon introducing interactions. It can be seen that $\epsilon < 1$ for this model for all values of $h$.}}
\label{Fig:nonint}
\end{figure}

\begin{figure}
\includegraphics[width=2.5in]{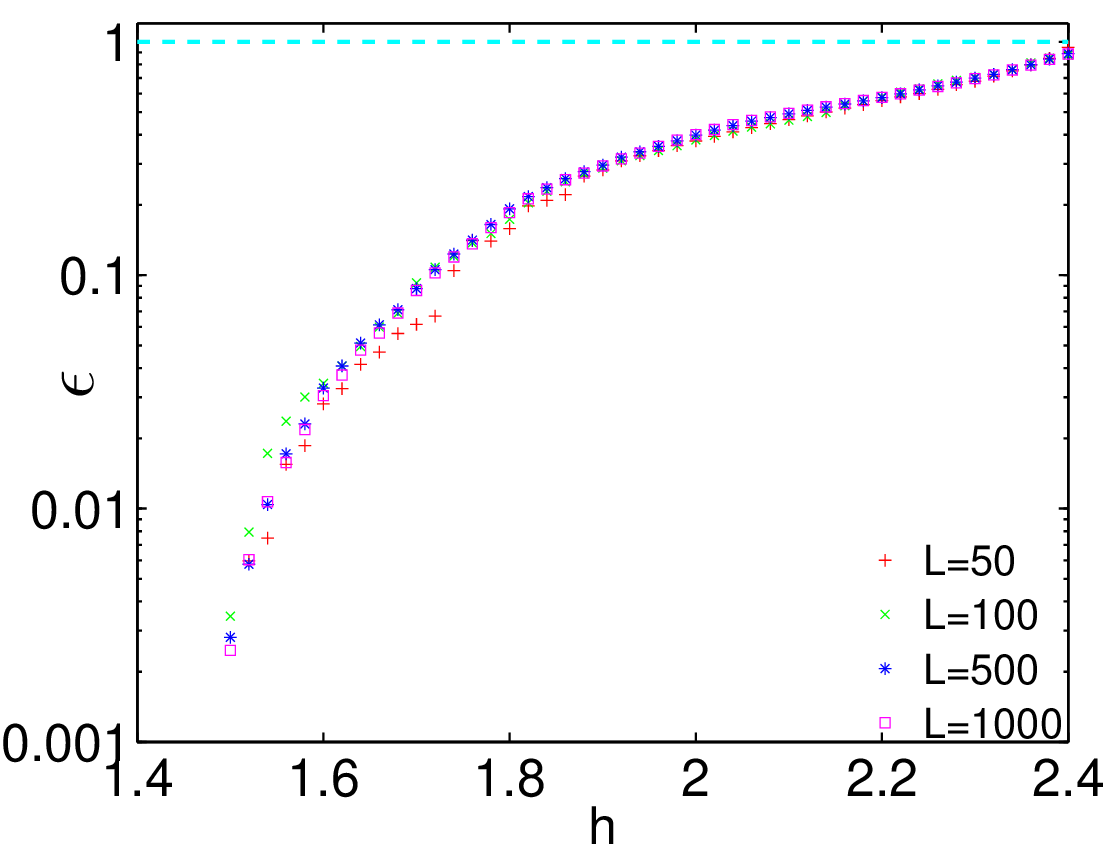}
\caption{\small{(Color Online) The variation of $\epsilon$ defined in Eqn.~\ref{Eq:defratio} with $h$ as obtained for $L=50,100,500,1000$  for model IV ( $t'=0.1$). The dashed line corresponds to $\epsilon=1$. Model IV always thermalizes upon introducing interactions. It can be seen that $\epsilon < 1$ for this model for all values of $h$.}}
\end{figure}

\section{Criterion for the occurrence of MBL}
\begin{table}
\begin{center}
\begin{tabular}{|c|c|c|c|p{2.5cm}|}
\hline
Model  & MBL phase & Thermal phase & $\nu$ &$\epsilon$\\
\hline
Model I & Yes & Yes & $<1$ & $>1$ (in MBL phase) and $<1$ (in thermal phase) \\
\hline
Model II & Yes & Yes & $<1$ & $>1$ (in MBL phase) and $<1$ (in thermal phase) \\
\hline
Model III & No & Yes & $1$ & $<1$ \\
\hline
Model IV & No &  Yes &  $<1$& $<1$\\
\hline
Model V & No & Yes & $>1$ & $<1$ \\
\hline
\end{tabular}
\end{center}
\caption{\small{A list of the models we study which shows whether they have thermal and MBL phases. The values of $\nu$ and $\epsilon$ for the different models obtained from exact diagonalization of systems of size $L=1000$. The values of $\nu$ agree with analytical results for the models for which they are available. The precise values of $\nu$ can depend on specific parameters of the different models but they are always bounded in the way shown in the above table.}}
\label{table:diff_exponents}
\end{table}
Table.~\ref{table:diff_exponents} summarizes the results of our calculations and the fifth column lists the values of $\epsilon$ defined in Eqn.~\ref{Eq:defratio}. The energy level spacing and entanglement entropy show that a thermal-MBL transition occurs in models I and II but not in models III, IV and V, which only have thermal phases.  At the MBL transition for models I and II, $\epsilon \sim 1$ as shown in Fig.~\ref{Fig:nonint} and in the MBL (thermal) phases $\epsilon > (<)1$.  On the other for model III
$\epsilon <1$ and this model always thermalizes. Similar values of $\epsilon (<1)$ are also obtained for models IV and V with $V=0$ which also always thermalize upon the introduction of interactions. Thus, our study  shows that the quantity $\epsilon$ can be used as a diagnostic to determine whether a system with a single particle mobility edge will display MBL upon the introduction of interactions. 

As mentioned earlier, it has been argued that the criterion for MBL to occur in a 1D system with a single particle mobility edge and a protected band of delocalized states upon introducing interactions is $\nu\ge 1$~\cite{nandkishore2014marginal}. It is thus interesting to ask whether a similar criterion applies even to the models we study with no protected delocalized states. We have calculated $\nu$ for these models for $V=0$ (listed in the fourth column of  table~\ref{table:diff_exponents}) using numerical exact diagonalization on systems up to $L=1000$ or known analytical results~\cite{dassarma.1990,modak.2015,de1998delocalization}.

The actual value of $\nu$ depends on the specific parameters of the model but is always bounded by or equal to 1 as indicated. It can be seen that both models I and II have $\nu < 1$ but show a thermal to MBL transition as a function of $h$ at a fixed value of the filling for fixed $V$. We have verified that such a transition holds down to values of $V$ as low as 0.2 below which finite-size effects become pronounced.  Significantly, $\nu$ is independent of $h$ (which is the parameter that is tuned to effect a the thermal-MBL transition in models I and II) in the models we have studied . It thus follows that there is no critical value $\nu_c$ such that models with $\nu > \nu_c$ exist in one phase and those with $\nu < \nu_c$ in the other upon introducing weak interactions. We note however that the model we study with $\nu > 1$ does not display MBL. Even though we have not been able to find a model with $\nu >1$ but $\epsilon > 1$ to study, we predict that should such a model exist it will display MBL upon the introduction of interactions.

\section{Non-interacting limit}
In the preceding sections, we have investigated the effect of interactions on the five models studied and proposed a criterion for the occurrence (or non-occurrence) of MBL in these models based on the properties of the single particle states of the corresponding non-interacting models. We now turn our attention to the many-body eigenstates of these models in the absence of interactions.

We ask whether the many-body eigenstates of these non-interacting models are localized or not in the sense of the entanglement entropy obeying an area law or volume law.
We also determine whether these states are `ergodic' or not. Note that the term ergodic here does not have its usual meaning in the sense of allowing thermalization for which interactions are required. Rather, we use the term ergodic to classify the strength of fluctuations of local operators, specifically when they do not increase with system size. We show that in the non-interacting models, there can exist transitions between such ergodic and non-ergodic phases characterized by the scaling with system size of the fluctuation of observables (specifically, the number of particles). Such a transition has been shown to exist in one of the models we study earlier ~\cite{Gan_non_int.2016}. Here, we investigate its occurrence in other models as well.  We also show that for states which are ergodic, the entanglement entropy of a state scales as the volume and has a value consistent with the thermodynamic entropy and the corresponding energy density.

Since the many-body eigenstates in the absence of interactions are simply obtained from populating single particle eigenstates (i.e. they are Slater determinants), we can diagonalize much larger systems (of size $L \sim 1000$) than in the case with interactions. 

\paragraph*{Entanglement entropy:}
In the non-interacting limit for Slater determinant states, the entanglement entropy can be calculated from the spectrum of the two point correlation function ~\cite{Peschel.2003}.
The $ij$th element of the two point correlation matrix is defined as
\begin{equation}
C_{ij}=\braket{\Psi|c_i^\dagger c_j| \Psi}
\end{equation}
where $\ket{\Psi}$ is a many body eigenstate and the indices $i$ and $j$ run over the subsystem whose entanglement with the rest of the system is calculated.
Diagonalizing the correlation matrix, the entanglement entropy (Renyi entropy) can be calculated from the eigenvalues of the correlation matrix.
\begin{equation}
S_2=- \sum_m log_2\left[ \zeta_m^2+(1-\zeta_m)^2\right]
\end{equation}
where $\zeta_m$ are the eigenvalues of correlation matrix C.

Once again many body eigenstates from the middle of the spectrum are considered. As such, mid-spectrum many body eigenstates of non interacting models are difficult to obtain for large system sizes as the Hilbert space dimension increases exponentially with system size and so does the number of steps required to sort the eigenvalues to identify the middle of the spectrum. However, since the many body density of states becomes sharply peaked at the middle of the spectrum with variance decreasing as $\frac{1}{\sqrt{L}}$ with increasing $L$, ~\cite{Gan_non_int.2016}, the mid-spectrum entropy can be calculated by averaging over the whole many body spectrum. This calculation can be simplified further by observing that most many-body eigenstates obtained by randomly occupying single particle states will lie in the middle of the spectrum. Thus, the mid-spectrum entropy can be obtained by averaging over a reasonably large number of randomly chosen many-body eigenstates. Here, we perform an average over $\sim 10^5$ states for the models we have studied for system sizes $L=100$, $200$ and $1000$.

\begin{figure}[t]
\includegraphics[height=2.5in, width=2.5in]{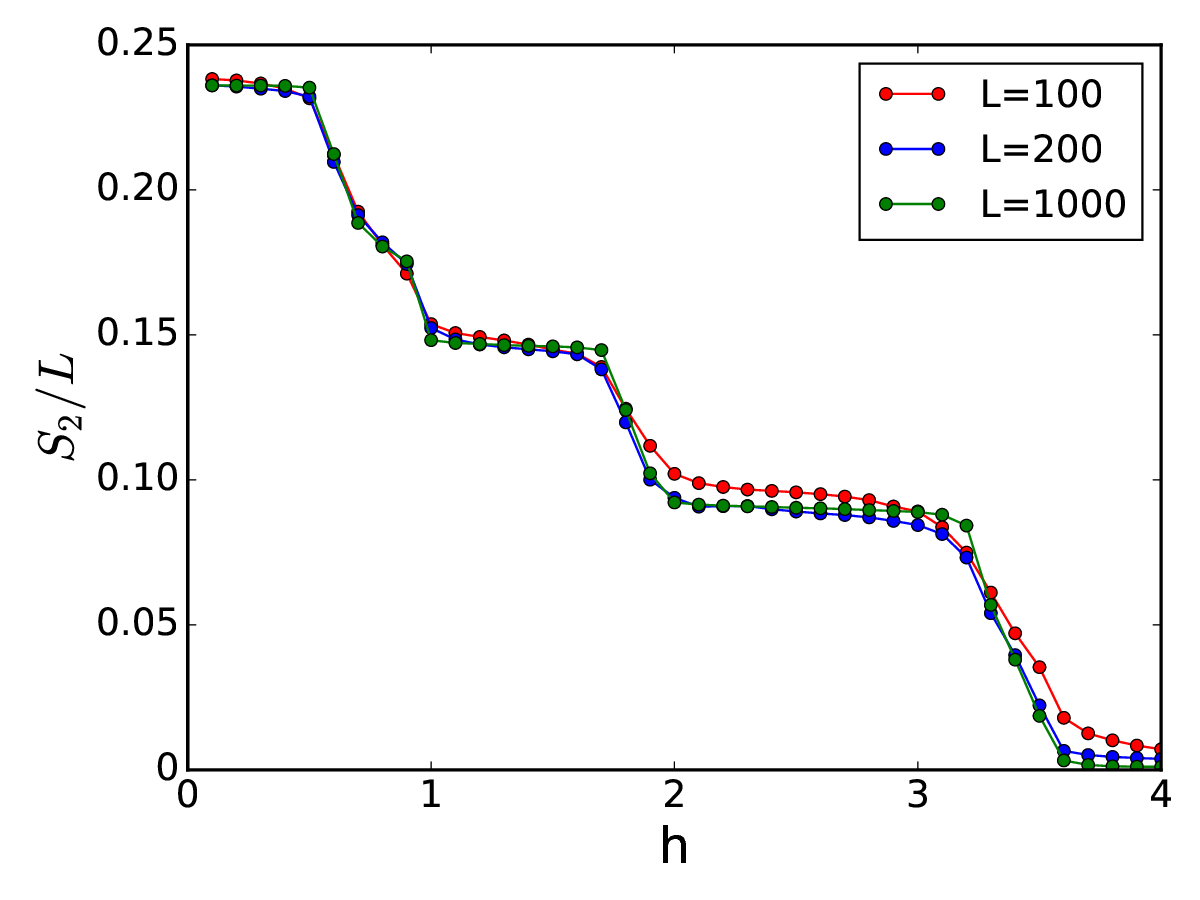} \\
\includegraphics[height=2.5in, width=2.5in]{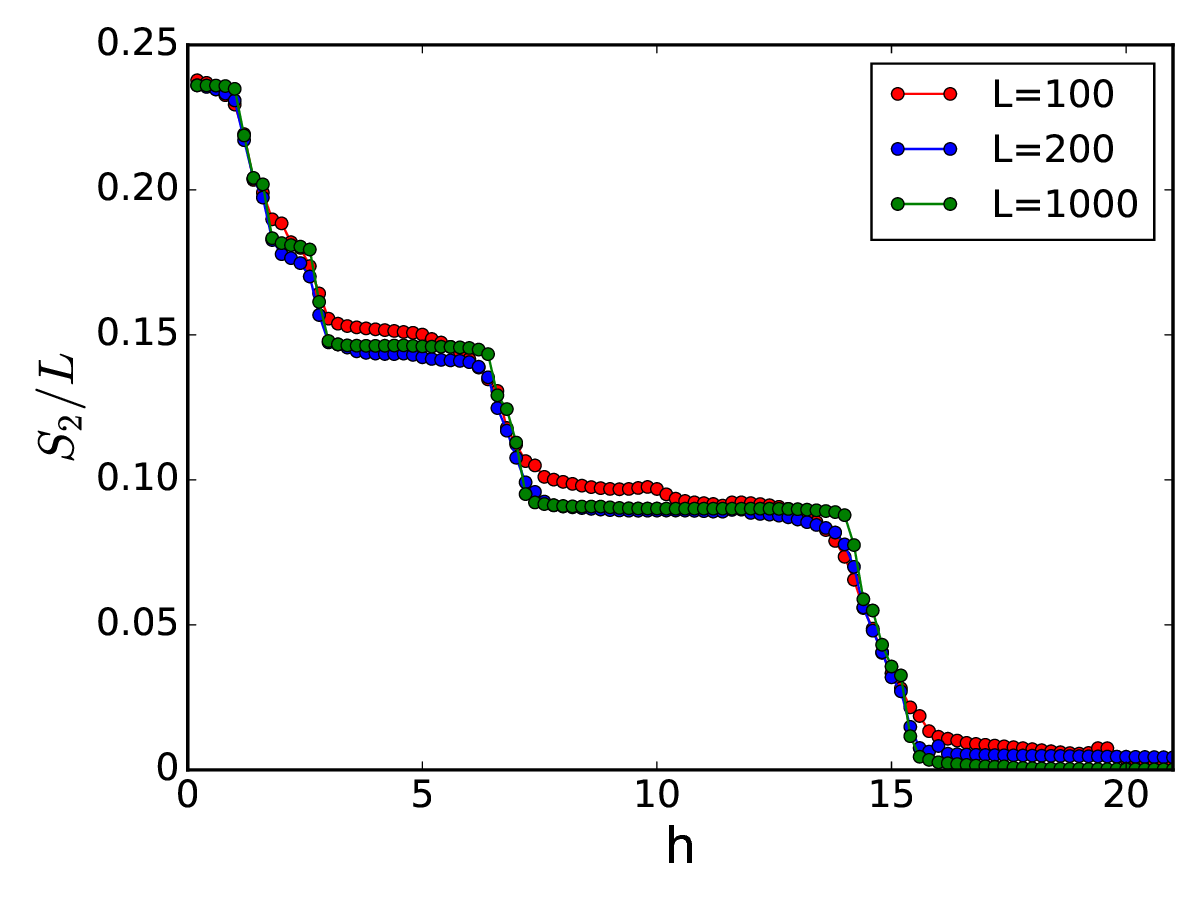}
\caption{\small{(Color Online) The variation of bipartite entanglement entropy $S_2/L$ with the onsite potential strength $h$ at half filling for (top) model-I ($\beta=-0.6$) and (bottom) model-II ($\beta=-0.75$) in the absence of interactions. The entanglement entropy is averaged over $10^5$ randomly chosen many body states. In model-I (model-II) for $h<0.5$ ($1.0$) the system is in the delocalized phase with the entropy consistent with its thermal value while for $h>3.6$ ($15.6$) the system is in the localized phase with negligible entanglement. In between these two distinct phases the system has two sub-thermal delocalized phases for $1.0t<h<1.7t$ and $2.0t<h<3.2t$ ($3.0<h<6.2$ and $7.4<h<14.0$) in model-I(II) where the entropy has sub-thermal values.}}
\label{Fig:ent-non-1}
\end{figure}

\begin{figure}[t]
\includegraphics[height=2.5in, width=2.5in]{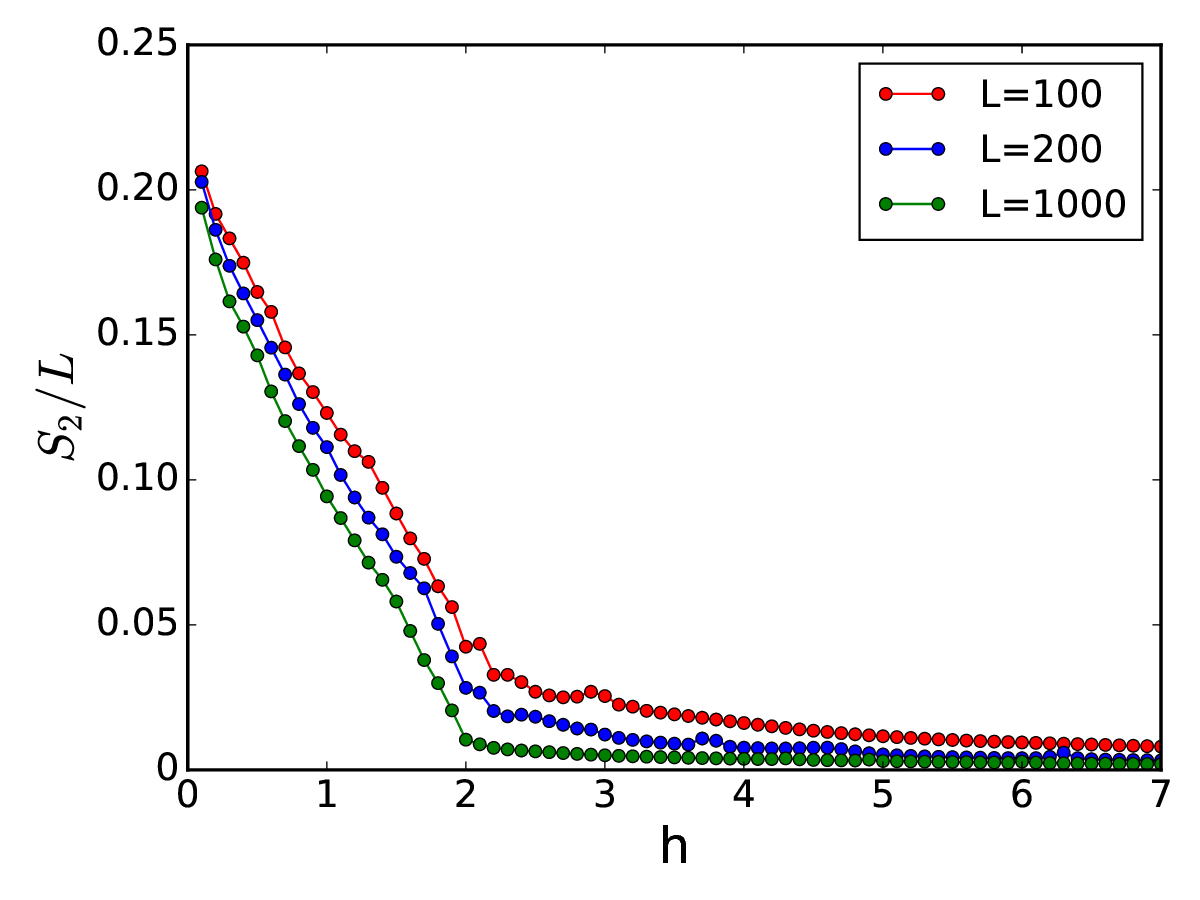} \\
\includegraphics[height=2.5in, width=2.5in]{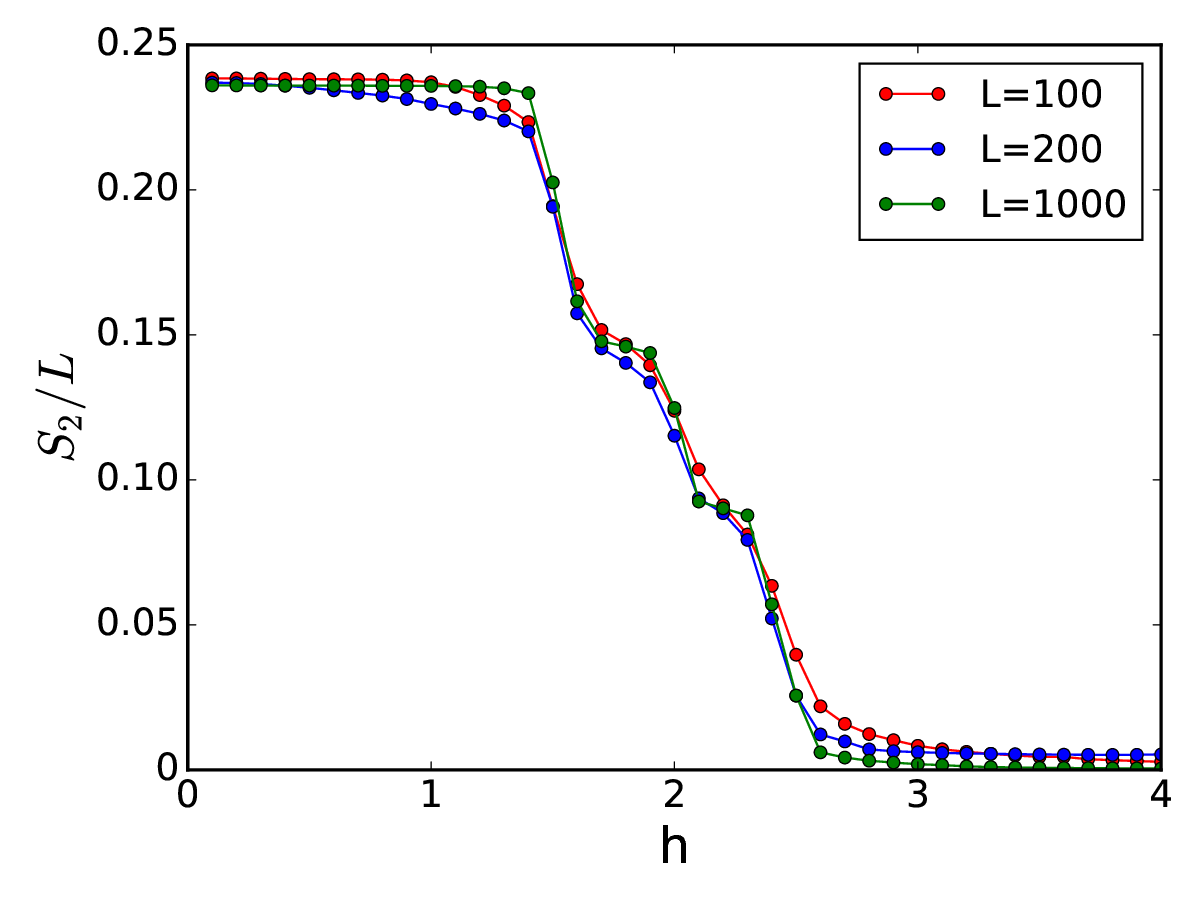}
\caption{\small{(Color Online) The variation of bipartite entanglement entropy $S_2/L$ with the onsite potential strength $h$ at half filling for (top) model-III ($n=0.5$), (bottom) model-IV ($t'=0.1$) in the absence of interactions. The entanglement entropy is averaged over $10^5$ randomly chosen many body states. In model-III for $h<2.0$ the system is delocalized but has sub-thermal values for the entropy while for $h>2.0$ the system is completely localized. A completely delocalized phase with entanglement entropy consistent with the thermal entropy and volume law scaling of entropy appears to be absent in this model and there is no crossover between the curves at different system sizes indicating the lack of any transitions.
In model-IV ,for $h<1.1$ the system is in the delocalized phase while for $h>2.6$ the system is in the localized phase. In between these two distinct phases, the system has two intermediate sub-thermal delocalized phases for $1.7<h<1.9$ and $2.1<h<2.3$.}}
\label{ent-non-2}
\label{Fig:nonint}
\end{figure}

As can be seen from Fig.~\ref{Fig:ent-non-1}, at small values of $h$ for models I and II, ($h<0.5$ for I and $h<1.0$ for II) the entanglement entropy scales with the subsystem size L/2 and is consistent with the value of the thermodynamic entropy indicating both delocalization and ergodicity. Note here that the entanglement entropy does not correspond to the thermal value as was the case in presence of interactions. It is due to the fact that we are calculating the infinite temperature entropy not from the eigenstates in the middle of the spectrum like in interacting calculations, instead we are averaging the entanglement entropy over the whole many body spectrum and the contribution from the states other than in the middle of the spectrum states reduce the value of the entanglement entropy ($\sim 0.25$) from its thermal value ($\sim 0.5$) even while retaining the volume law scaling~\cite{Gan_non_int.2016}.
 For large $h$ ( $h>3.6$ for I and $15.6$ for II) the entanglement entropy scales as $L^0$ (area law) indicating localization and thus also non-ergodicity. This is also the behavior observed in the Aubrey-Andre (AA) model~\cite{Gan_non_int.2016} which has a localization-delocalization transition as a function of the onsite potential $h$ but with no mobility edge. However, unlike in the AA model, there are intermediate phases in models I and II, where the entropy scales as a volume law but with a value which is sub-thermal which appear as plateaus in the Fig~\ref{Fig:ent-non-1}. This behavior was noted earlier only for model I with $\beta=-0.2$~\cite{Gan_non_int.2016}.  There are four such distinct phases at three critical values of $h$ corresponding to the transitions between the phases which can be determined by locating the crossovers between the entanglement entropy curves for different system sizes. The first transition is from the thermal delocalized phase to the first sub-thermal delocalized phase at $h=0.6$ for model I and $h=1.2$ for model II. The next transition occurs between the first and second subthermal delocalized phases at $h=1.8$ for model I and 
$h=6.8$ for model II. The final transition occurs between the second subthermal delocalized phase and the localized phase at  $h=3.3$ for model I and $h=14.8$ for model II. We will study ergodicity or lack thereof in these models from the point of view of particle number fluctuations in the next section.

We now turn to Fig.~\ref{ent-non-2} for models III and IV. Model III appears not to possess a thermal value of the entanglement entropy at any value of $h$. 
Model III appears not to possess an entanglement entropy consistent with the thermal value at any value of $h$. It also appears that the entanglement entropy scales slower than volume law ($2S_2/L$ decreases with the system size L) and the curves for different system sizes do not cross each other for any values of $h<2.0$. It should be noted that model III appears to always be thermal upon turning on weak interactions even as the non-interacting model does not seem to possess a delocalized ergodic phase at any value of $h$. It is possible that this is a finite-size effect and model III does not display behavior characteristic of the thermodynamic limit at the same system sizes as models I and II. However, currently accessible system sizes appear not to be sufficient to settle this issue.

Model IV displays plateaus in the entanglement entropy curve corresponding to intermediate sub thermal phases like models I and II. The three critical values of $h$ for the transitions between the four phases: ergodic phase, first delocalized sub thermal phase, second delocalized subthermal phase and localized phase are $1.5$, $2.0$, $2.4$ respectively. It should be noted that this model too like model III does appear to thermalize upon the introduction of weak interactions but the scaling of the entanglement entropy in the non-interacting limit appears to be quite distinct from that of model III.

\paragraph*{Particle number fluctuation:}
We have also calculated the fluctuation in the number of particles $N_{s}$ in one half of the system with a view towards further understanding the nature of ergodicity in the the different phases of the models identified from the calculation of the entanglement entropy. 
\begin{equation}
N_{s}=\sum_i \braket{\Psi | n_i |\Psi}
\end{equation}
where $i$ runs over all the lattice sites in the subsystem (one half of the whole system). The average of this quantity $\braket{N_s}$ is calculated by averaging over distinct mid-spectrum states. The fluctuation $\sigma_s$ in $N_s$ is defined as 
\begin{equation}
\sigma_{s}=\sqrt{\braket{N_{s}^2}-\braket{N_{s}}^2},
\end{equation} 
where again, the averaging is preformed over distinct mid-spectrum many-body eigenstates. Note that $\sigma_s$ is not the same as the fluctuation (square of the uncertainty) of $N_s$ in a single energy eigenstate.  The distinction between ergodicity or lack thereof in these non-interacting models arises from the fact that in the ergodic phase, $\sigma_s$ is independent of system size (intensive) going to a constant at large $L$  while in the non-ergodic phase, it scales as $\sqrt{L}$, or equivalently $\sqrt{\braket{N_s}}$ (extensive), since we work at fixed density~\cite{Gan_non_int.2016}. This notion of ergodicity is distinct from the one in the presence of interactions, where the number fluctuations scale to zero exponentially in the thermodynamic limit~\cite{deutsch1991}.

\begin{figure}[t]
\includegraphics[height=2.5in, width=2.5in]{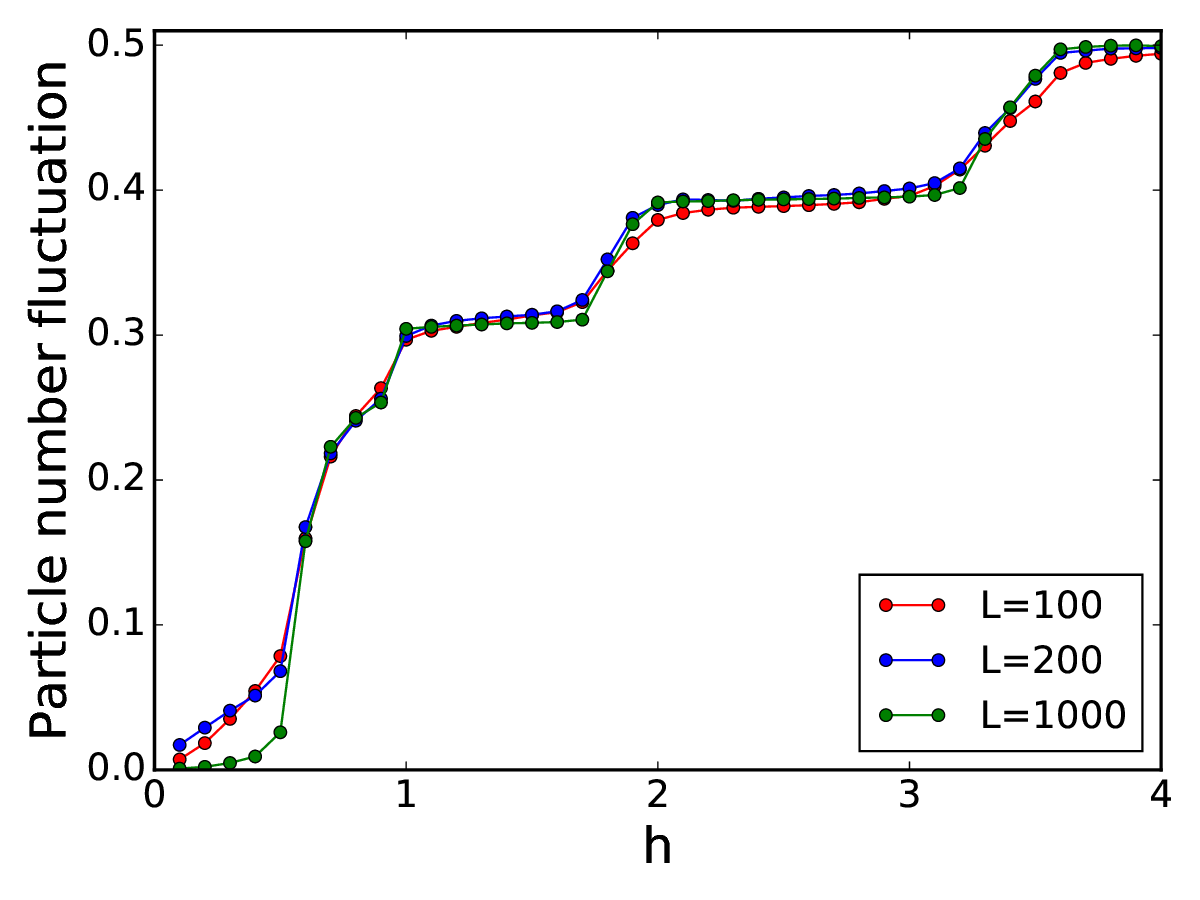} \\
\includegraphics[height=2.5in, width=2.5in]{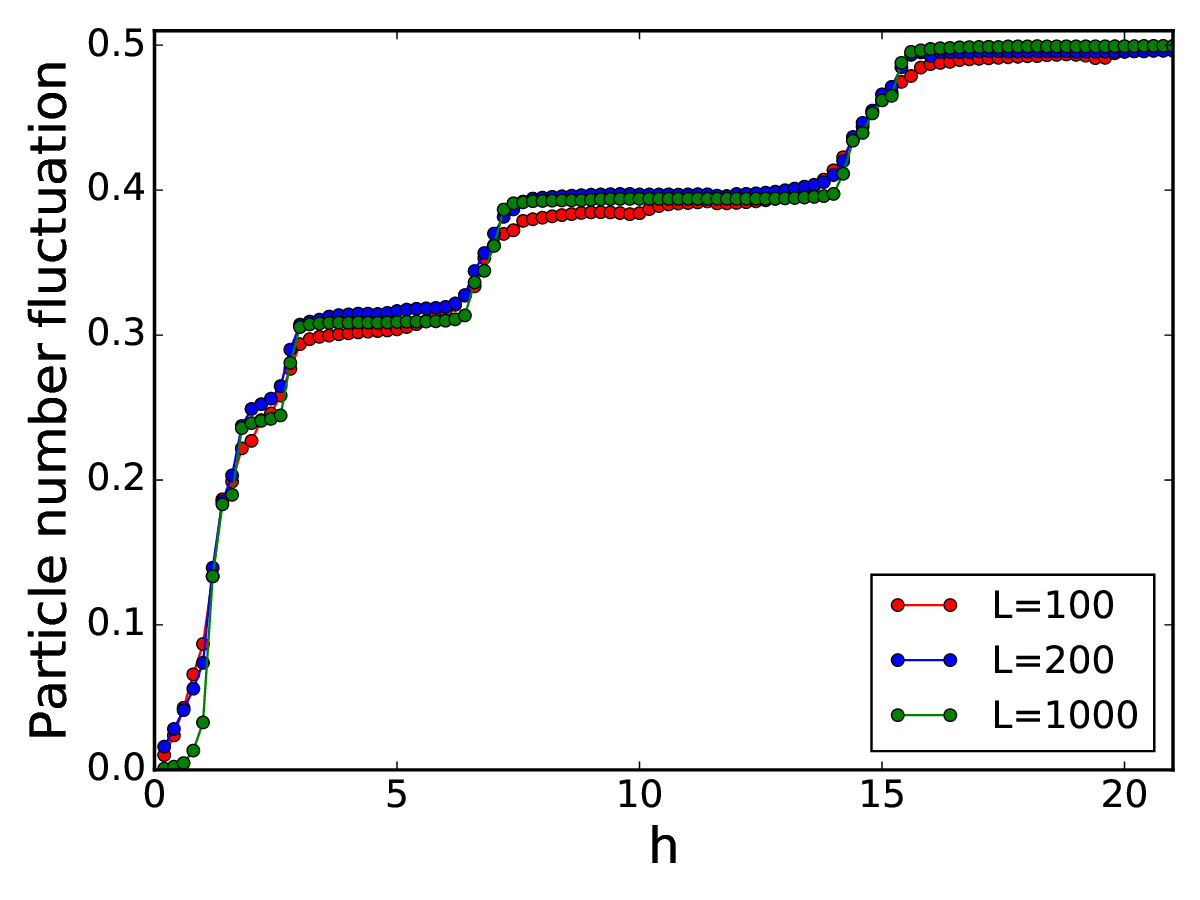}
\caption{\small{(Color Online) The variation of the particle number fluctuation ($\sigma_{sub}/\sqrt{\bar{N_s}}$) in one half of the system with the onsite potential strength $h$ at half filling for (top) model-I ($\beta=-0.6$), (bottom) model-II ($\beta=-0.75$) in the absence of interactions. In model-I (II) for $h<0.5$ ($1.0$) the system is in the thermal phase with intensive $\sigma_s$ while for $h>3.6$ ($15.6$) the system is in a non-thermal phase with $\sigma_s$ scaling as $\sqrt{N}$. In between these two distinct phases the system has two intermediate sub-thermal phases for $1.0<h<1.7$ and $2.0<h<3.2$ ($3.0<h<6.2$ and $7.4<h<14.0$) in model-I (II) where the fluctuation is smaller than the value in the non-thermal phase but is extensive quantity having $\sqrt{N}$ scaling.}}
\label{num-1}
\end{figure}
The particle number fluctuation plot (Fig.~\ref{num-1}) too displays multiple phases in models I and II.
For $h<0.5$ ($<1.5$), the particle number fluctuations are intensive indicating ergodicity while for $h>3.6$ ($>15.6$) they scale as $\sqrt{N}$ implying non-ergodicity. In between these two regimes plateaus in the plot indicate the presence of two intermediate phases in both models where $\sigma_{s}$ is smaller than its value in the large $h$ regime but still scales as $\sqrt{N}$ implying non-ergodicity. The critical values of $h$ at which the transitions to the intermediate phases occur match those obtained from the entanglement entropy (Fig.~\ref{Fig:ent-non-1}) thereby confirming that the intermediate phases are extensive (volume law scaling of the entanglement entropy) but non-ergodic (extensive scaling of particle number fluctuations). The small $h$ phase is extensive and ergodic and the large $h$ phase is localized (and thus also non-ergodic).
\begin{figure}[t]
\includegraphics[height=2.5in, width=2.5in]{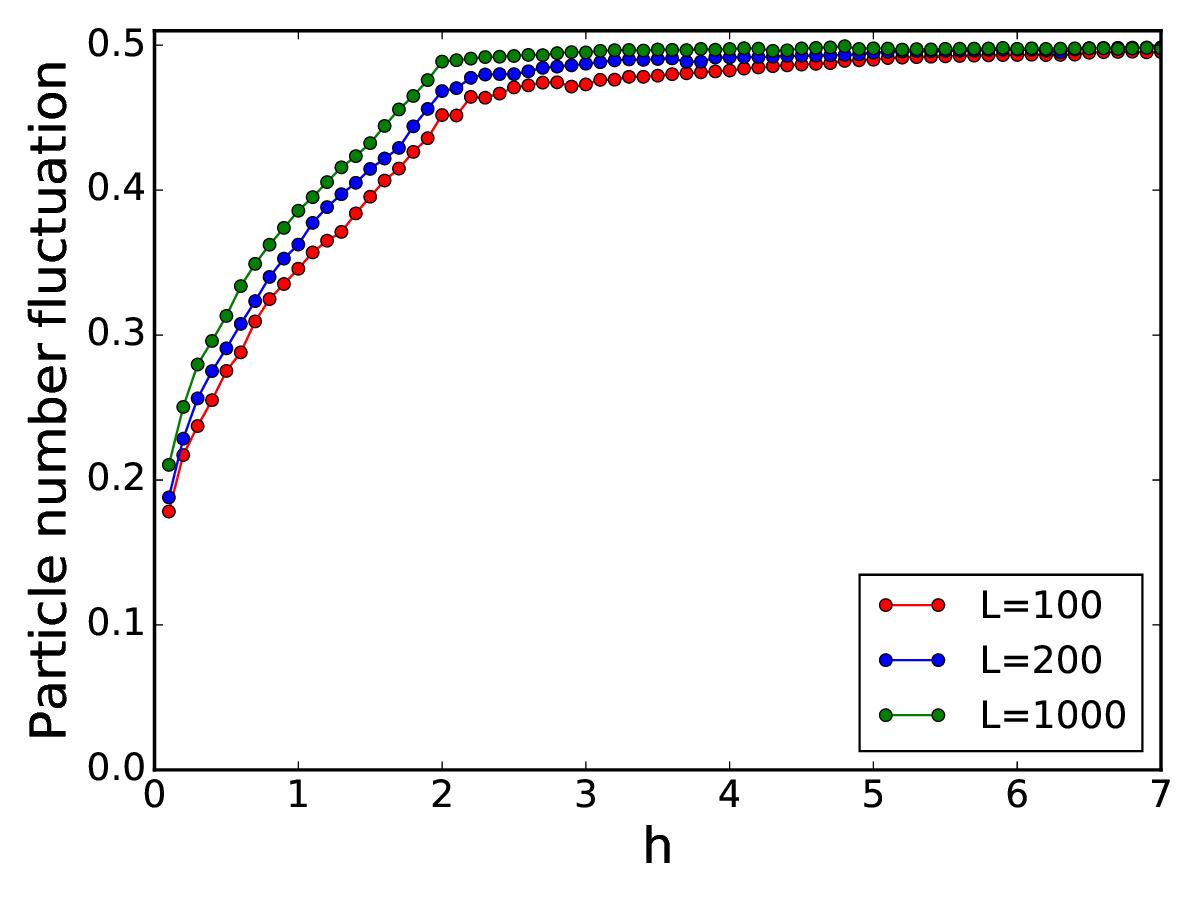} \\
\includegraphics[height=2.5in,width=2.5in]{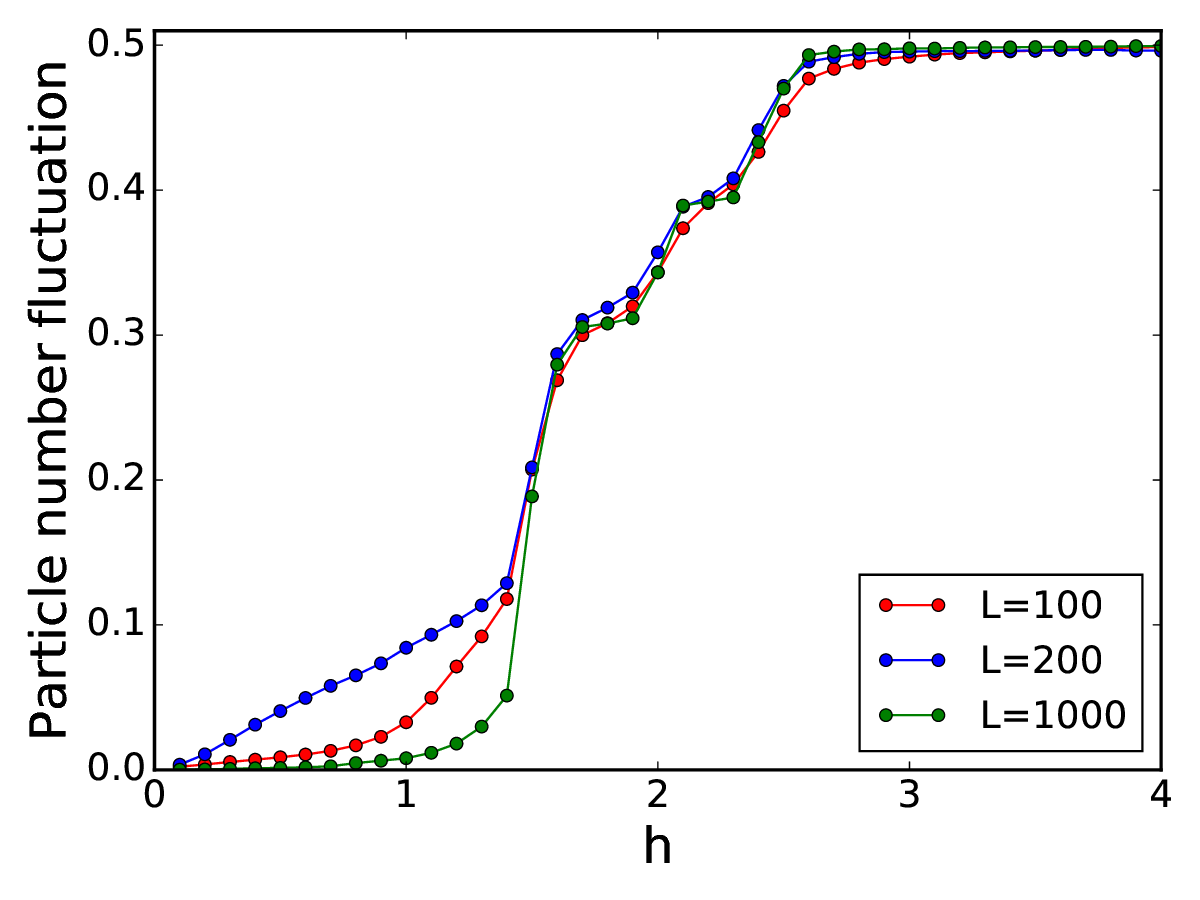}
\caption{\small{(Color Online) The variation of particle number fluctuation ($\sigma_s/\sqrt{\bar{N_s}}$) in one half of the system with the onsite potential strength $h$ at half filling for (top) model-III ($n=0.5$) and (bottom) model-IV ($t'=0.1$) in the absence of interactions. In model-III for $h<2.0$, $\sigma_s$ appears to scale as $N^\gamma$ where $\gamma>1/2$ which is indicative of neither ergodicity or localization. For $h>2.0$ the system shows extensive $\sigma_s$ with $\sqrt{N}$ scaling indicative of localization. The thermal phase with very small and intensive $\sigma_s$ is absent in this model and there is no crossover between the curves at different system sizes indicating the lack of any transitions. In model-IV for $h<1.1$, $\sigma_s$ is intensive and small in magnitude value indicating ergodicity while for $h>2.6$, $\sigma_s$ is extensive and $\sim 0.5$ corresponding to a non-ergodic phase. In between there are two plateaus corresponding to intermediate values of $\sigma_s$ which scales as $\sqrt{N}$}}
\label{num-2}
\end{figure}

In model-III as seen from Fig.~\ref{num-2} there appears to be no thermal phase with $\sigma_s$ scaling as an intensive quantity.
Instead for $h<2.0$, where the system has a single particle mobility edge, i.e., has both delocalized and localized single particle states, $\sigma_s \sim N^\gamma$ where $\gamma>1/2$. For $h>2.0$ the system has only localized states and exhibits non-ergodic behavior with $\sigma_s$ scaling as $\sqrt{N}$. $\sigma_s$ increases monotonically with $h$ for all values of $L$ studied indicating that the observed scaling $\sigma_s \sim N^\gamma$ with $\gamma>1/2$  can crossover to $\sigma_s \sim \sqrt{N}$ at even larger values of $L$. 
There are no plateaus for this model.

Model IV displays four different phase like models I and II. It has an ergodic phase for $h<1.1$, the first intermediate delocalized non-ergodic phase for $1.7<h<1.9$, second intermediate delocalized non-ergodic phase for $2.1<h<2.3$ and finally a localized phase for $h>2.6$.
As in the case of models I and II, the critical values of $h$ for the transition between these phases are the same as found from the entanglement entropy (Fig.~\ref{ent-non-2}).

\begin{figure}[t]
\includegraphics[height=2.5in, width=2.5in]{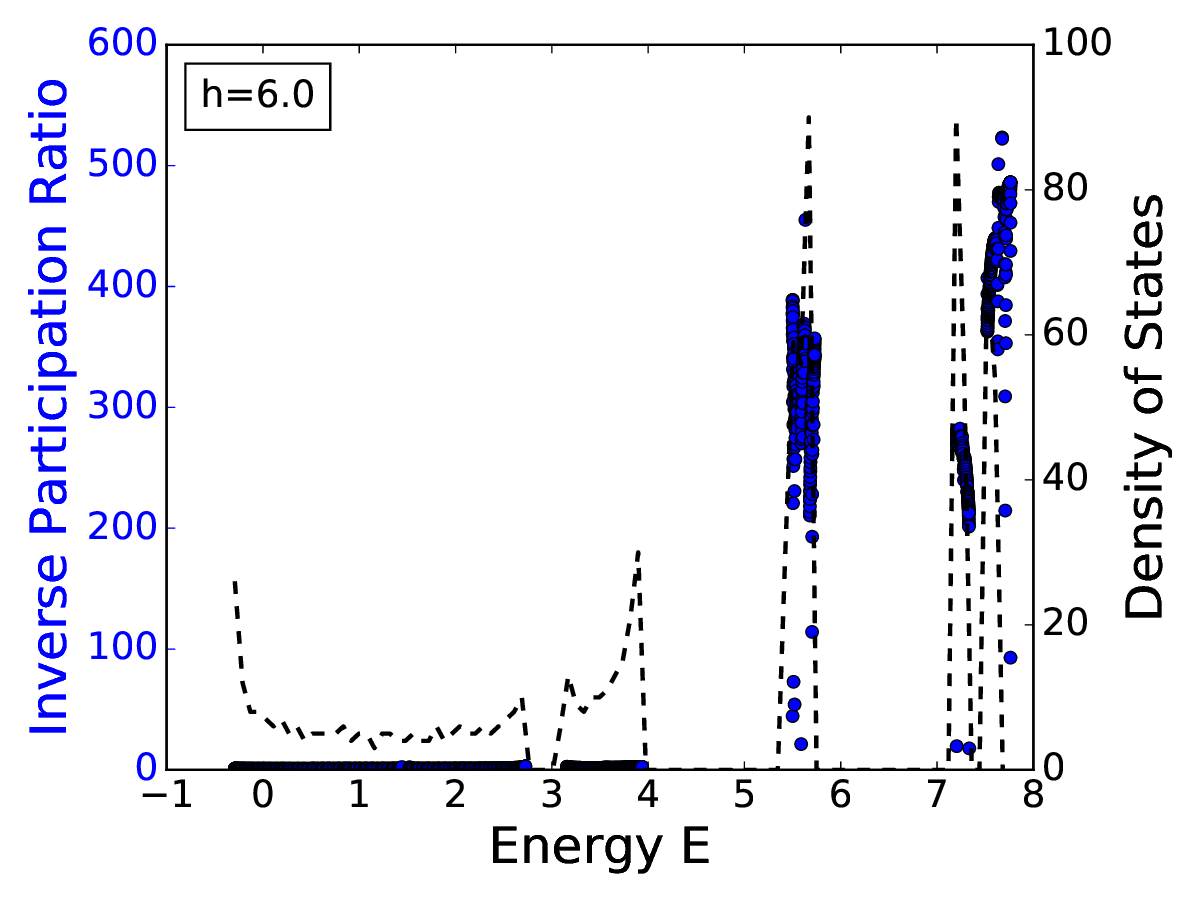}
\caption{\small{(Color Online) The density of states (DOS) and IPR of all the states of the single particle spectrum of model II with $\beta=-0.75$ at $h=6.0$. It can be seen that there are bands of states separated by gaps in the spectrum. Further a particular band contains either only localized or delocalized states. The higher energy bands are delocalized and the lower energy bands are localized} }
\label{Fig:Iprdos}
\end{figure}

The coexistence of localized and delocalized states in the single particle spectrum (i.e. the presence of a mobility edge) does not guarantee the existence of intermediate phases as as demonstrated by our study of model III. The difference in behavior of models I, II and IV on the one hand and model III on the other can be attributed to the the presence of bands in the spectra of models I, II and IV. As $h$ is increased, these bands get localized in succession. Whenever a particular band becomes completely localized while the next one is still delocalized, the entanglement entropy and the particle number fluctuation attain stable values giving rise to a delocalized non-ergodic phase. A small increase in $h$ localizes the next band resulting in an abrupt change in the values of entanglement entropy and particle number fluctuation producing a new plateau. This can be seen in Fig.~\ref{Fig:Iprdos} which shows the existence of bands in the spectrum of model II which are either localized or delocalized. The volume law scaling of the entropy arises from the existence of delocalized single particle states while non-ergodicity is due to the existence of localized states. At the two extremes, very small $h$ and very large $h$, there are no localized and delocalized bands respectively giving rise to ergodic and localized phases. The values of the entropy and fluctuation at plateau for a given model appears to only depend on the ratio of the number of delocalized to localized states.

It is tempting, by analogy, to compare the existence of plateaus in the intermediate phases to those in the quantum Hall effect, which are due to the presence of bands (Landau bands) that get filled successively upon increasing the magnetic field. However, there is no notion of quantized values at the plateaus here (and indeed the plateaus have different values for the different models) as opposed to the quantum Hall effect, where the values are quantized.

Model III does not possess bands like models I, II and IV and consequently does not display plateaus. 
However, it still possesses a mobility edge (and hence localized and delocalized) states but does not seem to 
display a proper delocalized phase, which could be a finite-size effect. 
The lack of bands implies that the ratio of the number of delocalized to localized states does not change 
in steps like models I, II and IV and instead changes gradually. Consequently, the entropy and fluctuation also do not exhibit steps.
Also note that model III is different from models I, II, and IV for at least the following two reasons, 1) the single particle delocalized states of model III are in the middle of the single particle spectrum (it has two mobility edges), and 2) the onsite potential of model III is very different from the onsite potentials of  model I, II, and IV (where it is `almost periodic'). These differences could potentially lead to the absence of plateaus and a proper delocalized-ergodic phase. However, further studies require in order to understand these features in more detail.

\begin{figure}[t]
\includegraphics[height=2.5in, width=2.5in]{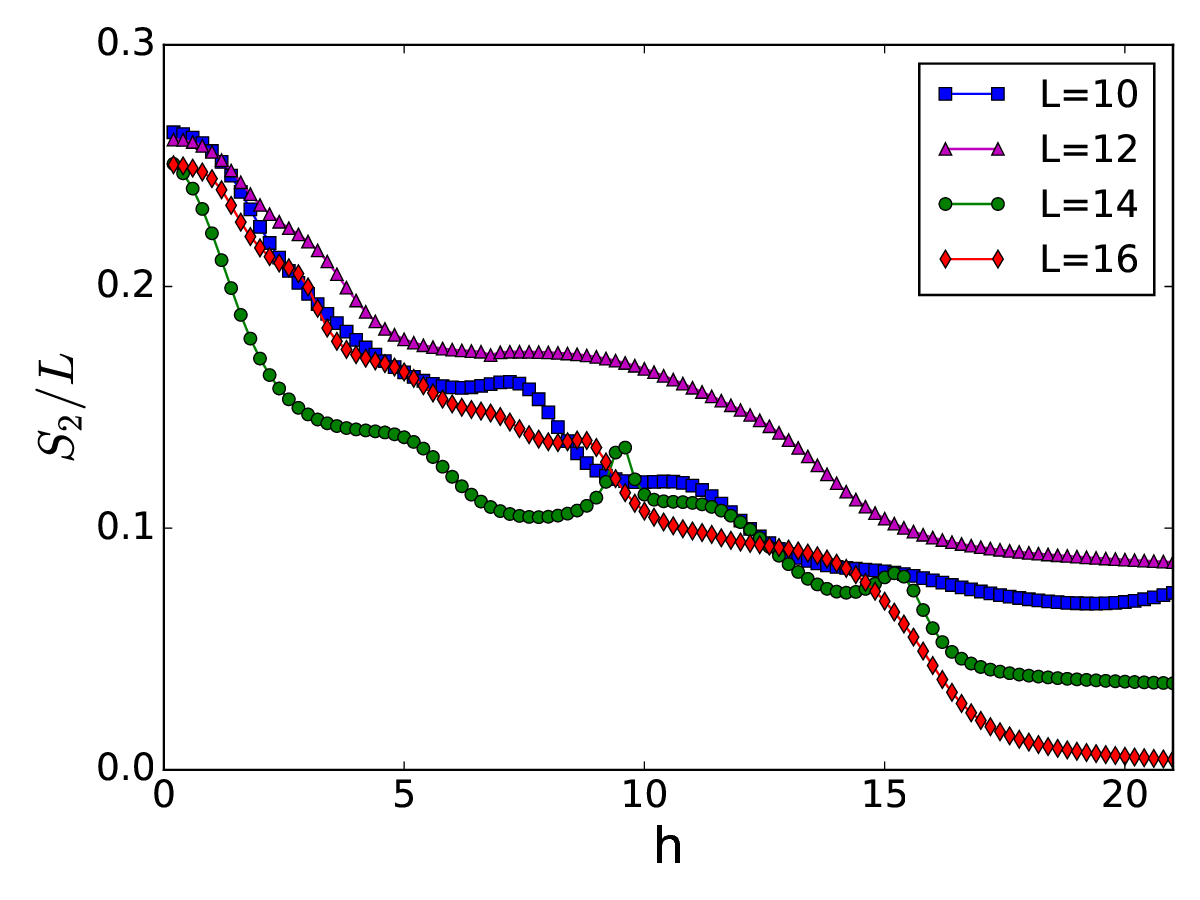}
\caption{\small{(Color Online) Variation of the bipartite entanglement entropy with $h$ for small system sizes $L=10-16$ in the non-interacting limit of model II with $\beta=-0.75$. It can be seen that in contrast to the bottom panel of Fig.~\ref{Fig:ent-non-1} with larger system sizes, there are no clear steps at these small system sizes. Thus, $L=10-16$ which are the typical sizes for studies on the interacting model might be too small to detect the intermediate phases if they exist in the presence of interactions.} }
\label{Fig:Smallsys}
\end{figure}

The criterion for the existence of MBL based on the parameter $\epsilon$ stated earlier can also be examined in the context of the phases of the non-interacting models. For models I and II, the critical value of $h$ predicted on the basis of the above criterion is very close to the critical value for the transition between the two intermediate phases. However, these intermediate phases appear not to exist in the presence of interactions. This indicates two possibilities: 1) The presence of weak interaction erases the intermediate phases from the system keeping only the thermal and MBL phases or 2) These intermediate phases exist in the interacting limit, but due to small system sizes $(L=16)$ the scaling of the entanglement entropy cannot distinguish between them. Evidence for the latter possibility comes from noting these phases are not visible and distinguishable even in the non-interacting limit for system sizes $L=10-16$ as can be seen from Fig.~\ref{Fig:Smallsys}.

For model-III the criterion based on $\epsilon$ predicts no thermal-MBL transition and numerically the system appears to be thermal in the presence of interactions. The non-interacting 
on the other hand appears not to have an ergodic phase. This is not necessarily contradictory since interactions generically have the tendency to cause thermalization. However, this fact requires further study to understand fully. 

Model IV also appears to thermalize upon the introduction of interactions in a manner consistent with the prediction based on $\epsilon$ as noted earlier. However, in the non-interacting limit, it too displays intermediate phases like models I and II but unlike them no thermal-MBL transition with weak interactions. This demonstrates that existence of transitions in the non-interacting limit between intermediate phases does not imply a thermal-MBL transition upon introducing interactions. 

The above considerations show that the criterion based on $\epsilon$ is a reliable predictor of the presence of MBL upon introducing interactions as opposed to the phases displayed by them in the non-interacting limit. Models I, II and IV display a similar pattern of ergodic, intermediate and localized phases in the non-interacting limit but upon introducing interactions, I and II display MBL while IV does not. On the other hand, III appears not to have an ergodic phase in the non-interacting limit but still thermalizes in the presence of interactions. All four of the models have single particle mobility edges.

\paragraph*{Dependence of Entanglement entropy on the number of localized states occupied:}
Finally, we examine the dependence of the entanglement entropy on the number of occupied localized states. To do this we have calculated the scaling of entanglement entropy with the sub-system size $\ell$ for different fractions of occupied localized single particle states. We have defined a parameter $f_{loc}$ which is the ratio of the occupied localized states to the total number of states occupied. In other words it is the fraction of the particles occupying localized single particle states. We have chosen the filling factor (i.e., ratio of the number of particles to the number of single particle states present in the system) to be 0.3 at $h=1.5$ in model I. The filling factor was chosen to allow for the possibility of all three types of many body states, 1) only localized single particle states occupied ($f_{loc}=1$), 2) only delocalized single particle states occupied ($f_{loc}=0$) and 3) both localized and delocalized single particle states occupied (($f_{loc}<1$). In each case, the entanglement entropy was averaged over $10^4$ many body eigenstates chosen by randomly populating single particle states keeping the total number of particles and $f_{loc}$  fixed. The entanglement entropy does not correspond to the infinite temperature (middle of the energy band) entropy for an arbitrary $f_{loc}$.

\begin{figure}[t]
\includegraphics[width=3.2 in]{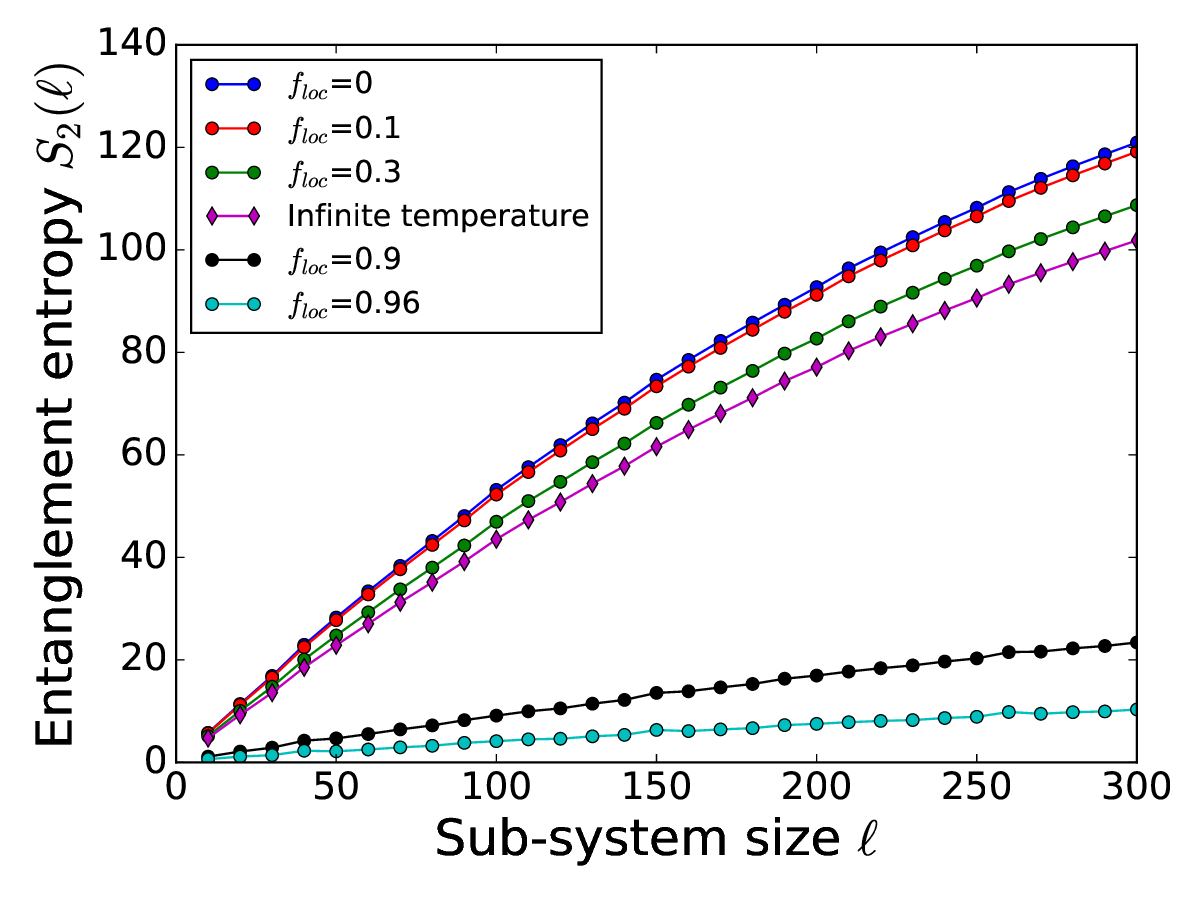}
\caption{Variation of entanglement entropy as a function of sub-system size $\ell$ for $h=1.5$ for model-I for $f_{loc}$=0,0.1,0.3,0.9,0.96. All the calculations are done in a system of size L=1000 with a filling factor 0.3. The entanglement entropy variation as a function of $\ell$ at infinite temperature is also plotted for comparison. The entanglement entropies are averaged over $10^4$ many body states in each case.} 
\label{sub}
\end{figure}

Fig.~\ref{sub} shows the variation of the entanglement entropy as a function of sub-system size $\ell$ for $h=1.5$ for model-I for $f_{loc}$=0,0.1,0.3,0.9,0.96. For comparison, the entropy for the infinite temperature case (for which the probability of occupying different single particle states is the same regardless of whether they are localized or delocalized~\cite{Gan_non_int.2016}), has also been plotted in the same figure. Fig.~\ref{sub} shows that the entanglement entropy for small values of $f_{loc}$ increases linearly with $\ell$ indicating a volume law with $f_{loc}$=0 corresponding to the largest value of the entanglement entropy showing the maximum amount of delocalization in the system. For large values of $f_{loc}$ (0.9,0.96) the entanglement entropy is almost independent of $\ell$ indicating area law entanglement and hence localization in the system.
Thus, we see that the value and scaling of the entanglement entropy depends on the fraction of localized states occupied in the single particle spectrum and changes from thermal to non-thermal behavior as the fraction of localized states increases.

\section{Conclusions}
We have investigated the effect of interactions on different models with mobility edges in the non-interacting limit using numerical exact diagonalization. We have demonstrated that MBL occurs in some of them (models I and II) but not in the others (models III, IV and V) and have proposed a criterion for whether MBL occurs in a model with a single particle mobility edge upon the introduction of interactions. The relevant quantity to calculate is $\epsilon$, the weighted ratio of participation ratios of the delocalized and localized states as given in Eqn.~\ref{Eq:defratio} and the criterion is that MBL occurs when $\epsilon > 1$ and the system thermalizes for $\epsilon < 1$. We have also shown that the presence or absence of MBL in models with unprotected delocalized states does not appear to depend on the value of the localization length exponent $\nu$. Further, we have studied entanglement entropy scaling and particle number fluctuations in models with single particle mobility edges in the non-interacting limit. These quantities can be used to detect the presence of localization and ergodicity in these models respectively. We have found the existence of non-ergodic delocalized phases intermediate to ergodic and localized phases in some of these models. However, we have also demonstrated that the presence or absence of these phases does not seem to be a predictor for the occurrence of MBL upon switching on weak interactions. We thus conclude that the  criterion based on the quantity $\epsilon$ introduced here is the most appropriate one to predict whether MBL occurs in models with single particle mobility edges upon introducing interactions.

\paragraph*{Acknowledgments:} RM acknowledges support from the UGC-BSR Fellowship and SM from the DST, Govt. of India and the UGC-ISF Indo-Israeli joint research program for funding.
   
%

\end{document}